\documentstyle[aps,preprint,tighten,epsf]{revtex}

\begin{document}

\preprint{}
\draft

\title{
Crossover between Luttinger and Fermi liquid behavior in weakly
coupled metallic chains }
\author{Peter Kopietz, Volker Meden\thanks{Present address: Department of 
Physics, Indiana University, Bloomington, IN 47405, U.S.A.},
and Kurt Sch\"{o}nhammer}
\address{
Institut f\"{u}r Theoretische Physik der Universit\"{a}t G\"{o}ttingen,\\
Bunsenstr.9, D-37073 G\"{o}ttingen, Germany}
\date{December 19, 1996}
\maketitle
\begin{abstract}
We use higher-dimensional bosonization to
study the normal state of electrons in weakly coupled metallic chains
interacting with long-range Coulomb forces.
Particular attention is paid to the 
crossover between Luttinger and Fermi liquid behavior as 
the interchain hopping $t_{\bot}$ 
is varied.
Although in the physically interesting case 
of finite but small $t_{\bot}$
the quasi-particle residue does not vanish, 
the single-particle Green's function  
exhibits the signature of Luttinger liquid behavior
(i.e. anomalous scaling and spin-charge separation)
in a large intermediate parameter regime.
Using realistic parameters, we find that the scaling behavior in this regime is
characterized by an anomalous dimension of the order of unity,
as suggested by recent experiments on quasi-one-dimensional conductors. 
Our calculation gives also new insights into the
approximations inherent in higher-dimensional bosonization; in 
particular, we show that the replacement of a curved 
Fermi surface by a {\it{finite}} number $M$ of flat patches can give
rise to unphysical nesting singularities in the single-particle
Green's function, which disappear only in the limit $M \rightarrow \infty$ or
if curvature effects are included.
We also compare our approach with other methods.
This work extents our recent letter
[Phys.\ Rev.\ Lett.\ {\bf{74}}, 2997 (1995).].

\end{abstract}
\narrowtext

\section{Introduction}

The properties of the normal metallic state of correlated electrons in
$d=1$ dimension are 
usually summarized under the name Luttinger liquid behavior \cite{Haldane81}.
The single-particle Green's function of a Luttinger liquid 
exhibits several striking differences to the 
Green's function of a conventional Fermi liquid:
the absence of a coherent quasi-particle peak, spin-charge separation, and 
anomalous scaling properties  characterized by interaction-dependent power laws.
Non-Fermi liquid behavior has recently been observed in several experiments.
First of all, the normal-state properties of the
high-temperature superconductors cannot be interpreted in terms
of a conventional Fermi liquid picture \cite{Varma89}. 
Motivated by this experimental fact,
Anderson \cite{Anderson92} 
proposed that even in $d=2$ the single-particle Green's functions of 
correlated electrons can exhibit Luttinger liquid behavior.
This unconventional normal-state behavior was then shown to be 
a possible key to understand the novel superconducting state.
In fact, the interlayer-tunneling theory of high-temperature
superconductivity advanced by Chakravarty, Anderson and 
co-workers \cite{Chakravarty94} is based 
on the fundamental assumption that
in the normal state the single-particle Green's function
$G ( {\bf{k}} , \omega )$ satisfies
for  wave-vectors ${\bf{k}}$ sufficiently close to the Fermi surface
and for sufficiently low frequencies $\omega$ (measured relative to the chemical 
potential) an anomalous scaling law 
of the form
 \begin{equation}
 G ( {\bf{k}}^{\alpha} + s {\bf{q}} , s \omega ) = s^{\gamma -1 }
 G ( {\bf{k}}^{\alpha} + { \bf{q}} , \omega )
 \label{eq:anomalscale}
 \; \; \; ,
 \end{equation}
where ${\bf{k}}^{\alpha}$ is a wave-vector on the Fermi surface, and 
$\gamma > 0$ is some interaction-dependent exponent, the anomalous dimension.
A finite value of $\gamma$ is one of the fundamental
characteristics of Luttinger liquid behavior, whereas in 
a Fermi liquid $\gamma = 0$.
It should be stressed, however, that a generally accepted microscopic
derivation of Eq.\ (\ref{eq:anomalscale}) for interacting
fermions in $d > 1$ does not exist.

Another class of materials where non-Fermi liquid behavior 
appears to have been
observed experimentally consists of weakly coupled
metallic chains \cite{Jerome91}, which are based on 
highly anisotropic conductors.
Although at sufficiently low temperatures $T$ 
these systems have the tendency to develop various 
types of long-range order,
above the ordering temperature
the metallic state displays clearly non-Fermi 
liquid behavior.
The interpretation of recent photoemission studies of these systems
in terms of the Luttinger liquid picture leads to 
values for the anomalous dimension $\gamma$ in the range
$1.0 \pm 0.2$ \cite{Dardel93,Nakamura94,Claessen95}.
Because in one-dimensional lattice models with short-range interactions
(such as the Hubbard model) the anomalous dimension 
is always small compared with unity \cite{Schulz90},
the large experimentally measured value of $\gamma$ suggests that
the long-range nature of the Coulomb interaction
must play an important role in these systems.
Of course, the experimental systems are not strictly one-dimensional,
because the interchain hopping $t_{\bot}$ is small but finite,
and even for $t_{\bot} = 0$ the 
electrons on different chains interact with
three-dimensional Coulomb forces.
We thus arrive at the problem of weakly coupled metallic chains,
which is the central topic of this paper.

In the last years the problem of coupled Luttinger liquids has been
studied by many 
authors \cite{Gorkov74,Lee77,Wen90,Bourbonnais91,Schulz91,Fabrizio92,Kusmartsev92,%
Castellani92,Yakovenko92,Finkelshtein93,Boies95,KMS95,Clarke96,Tsvelik96}. 
Given the fact that in the absence of interchain hopping the
non-perturbative bosonization approach can be used to 
calculate the single-particle Green's function in a controlled way,
some authors \cite{Wen90,Bourbonnais91,Castellani92,Boies95,Clarke96,Tsvelik96} 
have attempted to supplement the
one-dimensional non-perturbative bosonization solution by some kind of
perturbation theory in powers of $t_{\bot}$. 
One disadvantage of this strategy is that
in this way interchain hopping and electron-electron interactions
between the chains are not taken into account on equal footing.
In this work we shall therefore adopt a different strategy:
Instead of combining one-dimensional bosonization with
perturbation theory in powers of $t_{\bot}$, we shall use
the recently developed higher-dimensional 
generalization of the bosonization 
method \cite{Luther79,Haldane92,Houghton93,Frohlich95,Kopietz94,Kopietzhab,Kopietz96} 
to treat interchain hopping and
interactions between the chains on equal footing.
Thus, our approach is non-perturbative in $t_{\bot}$ and in the interaction.
In particular, for $t_{\bot} = 0$ we recover the well-known
one-dimensional bosonization result for the Green's function of the
Tomonaga-Luttinger model.

The rest of this paper is organized as follows:
In Sec.\ \ref{sec:bos} we briefly describe the higher-dimensional
bosonization approach. In particular, we emphasize 
the approximations inherent in this
approach and discuss its limitations.
In Sec.\ \ref{sec:thecoulomb} we shall show how the
problem of electrons that are confined to one-dimensional chains 
(i.e. without interchain hopping)
and interact with three-dimensional Coulomb forces
can be solved in a straightforward way with higher-dimensional bosonization.
Of course, this problem can also be solved by means of the conventional
one-dimensional bosonization technique \cite{Solyom,Schulz83,Penc93}. 
However, as shown
in Sec.\ \ref{sec:finiteinter}, our approach 
can also handle the case of finite
interchain hopping $t_{\bot}$ without having to rely on
an expansion in powers of $t_{\bot}$ or in the interaction.
Although the quasi-particle residue is found to be finite
for any $t_{\bot } \neq 0$, we show by explicit calculation 
that there exists a large intermediate regime where
the equal time Green's function satisfies an anomalous scaling relation
similar to a Luttinger liquid. In Sec.\ \ref{sec:wenapproximation} we compare
our results with those obtained by other approaches.
Finally, in Sec.\ \ref{sec:conclusions} we present our conclusions. 

\section{Higher-dimensional bosonization}
\label{sec:bos}

In this section we shall give a brief summary of
the higher-dimensional bosonization approach. 
The foundations of this novel non-perturbative
approach to the fermionic many-body problem go 
back to ideas due to Luther \cite{Luther79}
and Haldane \cite{Haldane92}. More recently, a number
of authors have applied this technique to
problems of physical interest \cite{Houghton93,Frohlich95,Kopietz94,Kopietzhab}.
The basis of higher-dimensional bosonization is the
subdivision of the Fermi surface into a finite number 
of patches $P^{\alpha}_{\Lambda}$, $\alpha = 1 , \ldots , M$. 
To be specific, let us consider in Fig.\ \ref{fig:quasi1d}  
a typical Fermi surface associated with a periodic array of chains
coupled by weak interchain hopping.
A possible subdivision into $M = 8$ 
patches is shown in Fig.\ \ref{fig:quasi1dpatch}.
Each patch $P^{\alpha}_{\Lambda}$ is then extended
into a $3$-dimensional squat box \cite{Houghton93}
as shown in Fig.\ \ref{fig:squat}.
The characteristic size $\Lambda$ of the patches should be chosen
sufficiently small so that 
within a given patch the curvature of the Fermi surface can be
locally ignored. On the other hand, the cutoffs $\Lambda$ and $\lambda$ cannot 
be taken arbitrarily small, because 
higher-dimensional bosonization becomes only useful 
in practice if it is possible
to ignore momentum transfer between different 
patches (the so-called around-the-corner processes \cite{Kopietz94}).
Of course, this is only possible if the interaction is dominated by forward scattering.
More quantitatively, we assume that the two-body interaction
$f_{\bf{q}}$ becomes negligibly small if $|{\bf{q}}|$ is larger than some
cutoff $q_c  \ll \mbox{min} \{ \Lambda , \lambda \}$.
The geometry is shown in Fig.\ \ref{fig:squat}.
Assuming that the nature of the interaction is such that the above condition
can be satisfied, it is reasonable
to make the following two approximations \cite{Kopietz94,Kopietzhab}:
First of all, we ignore momentum transfer
between different boxes (diagonal-patch approximation). 
The relative number of matrix elements neglected in this case is in $d$ dimensions of order
$q_c^d / ( \Lambda^{d-1} \lambda ) \ll 1$.
The second fundamental approximation is the local
linearization of the energy dispersion.
Measuring wave-vectors locally with respect to coordinate systems
located at points ${\bf{k}}^{\alpha}$ in the centers of the patches
(see Fig.\ \ref{fig:squat}), a general energy dispersion 
$\epsilon_{\bf{k}}$ 
may be expanded as
 \begin{equation}
 \epsilon_{ {\bf{k}}^{\alpha} + {\bf{q}} } =
 \epsilon_{\bf{k}^{\alpha} } + \xi^{\alpha}_{\bf{q}}
 \; \; \; , \; \; \; 
 \xi^{\alpha}_{\bf{q}} =
 {\bf{v}}^{\alpha} \cdot {\bf{q}} +
 \frac{  (q_{\|}^{\alpha})^2 }{2 m^{\alpha}_{\|}  }
 +
 \frac{ (\bf{q}_{\bot}^{\alpha})^2}{2 m^{\alpha}_{\bot}  }
 + \ldots 
 \; \; \; ,
 \label{eq:energy}
 \end{equation}
where $q_{\|}^{\alpha} = {\bf{q}} \cdot \hat{\bf{v}}^{\alpha}$ 
and ${\bf{q}}_{\bot}^{\alpha} = {\bf{q}}
- ( {\bf{q}} \cdot \hat{\bf{v}}^{\alpha} )
 \hat{\bf{v}}^{\alpha}$. Here
$\hat{\bf{v}}^{\alpha}  = {\bf{v}}^{\alpha} /  |{\bf{v}}^{\alpha} |$  
is a unit vector in the direction
of ${\bf{v}}^{\alpha}$.
Note that in general 
the masses $m^{\alpha}_{\|}$ and $m^{\alpha}_{\bot}$ depend on the patch-index.
Because in the grand-canonical formulation of statistical mechanics 
the energy dispersion appears only in the combination
$\epsilon_{\bf{k}} - \mu$,
the chemical potential $\mu$ cancels
the constant $\epsilon_{\bf{k}^{\alpha}}$ in Eq.\ (\ref{eq:energy})
provided ${\bf{k}}^{\alpha}$ is chosen to lie exactly on the Fermi surface.
We now ignore all terms in Eq.\ (\ref{eq:energy}) that are quadratic and higher
order in ${\bf{q}}$, i.e.\ we approximate
$\xi^{\alpha}_{\bf{q}} \approx {\bf{v}}^{\alpha} \cdot {\bf{q}}$.
In particular, the term $({\bf{q}}^{\alpha}_{\bot})^2 / ( 2 m_{\bot}^{\alpha} )$ is neglected.
Note that retaining this term 
would lead to patches with non-zero curvature.
Once we accept the validity of the above approximations, the
single-particle Green's function can be calculated without further approximation
in arbitrary dimension. 
As shown in Refs.\ \cite{Kopietz94} and \cite{Kopietzhab},
the result for the Matsubara Green's function at finite
inverse temperature $\beta$ can be written as
 \begin{equation}
 G ( {\bf{k}}^{\alpha} + {\bf{q}} , i \tilde{\omega}_{n} )
 =
 \int d {\bf{r}} \int_{0}^{\beta} d \tau 
 e^{ - i (  {\bf{q}}  \cdot  {\bf{r}}
 - \tilde{\omega}_{n}  \tau  ) }
 G^{\alpha}_0 ( {\bf{r}} , \tau )
 e^{Q^{\alpha} ( {\bf{r}} , \tau )}
 \label{eq:Gbosres}
 \; \; \; ,
 \end{equation}
where 
 \begin{equation}
 {{G}}^{\alpha}_{0} ( {\bf{r}}  , \tau  )
  = 
 \frac{1}{\beta V} \sum_{ {\bf{q} } , \tilde{\omega}_n  } 
 \frac{ e^{ i ( {\bf{q}} \cdot {\bf{r}} - \tilde{\omega}_{n} \tau )}}{ i \tilde{\omega}_{n}
 -  {\bf{v}}^{\alpha} \cdot {\bf{q}} } 
 \; \; \; ,
 \label{eq:G0res}
 \end{equation}
and the Debye-Waller factor $Q^{\alpha} ( {\bf{r}} ,\tau )$
is given by
 \begin{equation}
 Q^{\alpha}
 ( {\bf{r}} , \tau )  = 
 R^{\alpha} 
 - S^{\alpha} ( {\bf{r}} , \tau )
 \;  \; \; ,
 \label{eq:Qlondef}
 \end{equation}
with
 \begin{eqnarray}
 R^{\alpha} 
  &  =  & 
 \frac{1}{\beta {{V}}} \sum_{ {\bf{q}} , \omega_m }  
 \frac{   f^{{RPA}} ( {\bf{q}} , i \omega_m ) }{
 ( i \omega_{m} - {\bf{v}}^{\alpha} \cdot {\bf{q}} )^{2 }}
   =  S^{\alpha} ( 0 , 0 ) 
  \label{eq:Rlondef}
 \; \; \; ,
 \\
 S^{\alpha} 
 ( {\bf{r}}  , \tau   ) 
  &  =  &
 \frac{1}{\beta {{V}}} \sum_{ {\bf{q}} , \omega_m }  
 \frac{   f^{{{RPA}}} ( {\bf{q}} , i \omega_m)
  \cos ( {\bf{q}} \cdot  {\bf{r}} 
  - {\omega}_{m}  \tau  ) 
 }
 {
 ( i \omega_{m} - {\bf{v}}^{\alpha} \cdot {\bf{q}} )^{2 }}
  \label{eq:Slondef}
 \; \; \; .
 \end{eqnarray}
Here $\tilde{\omega}_n = 2 \pi (n + \frac{1}{2} ) / \beta $ are fermionic
Matsubara frequencies, and $\omega_m = 2 \pi m / \beta $ are bosonic ones.
$f^{{{RPA}}} ( {\bf{q}} , i \omega_m)$ 
is the  screened interaction in random-phase approximation (RPA), which
is given in terms of the bare interaction $f_{\bf{q}}$ and the
non-interacting polarization $\Pi_0 ( {\bf{q}} , i \omega_m )$ via
the usual relation
 \begin{equation}
f^{{{RPA}}} ( {\bf{q}} , i \omega_m) = \frac{ f_{\bf{q}} }{ 1
+ f_{\bf{q}} \Pi_0 ( {\bf{q}} , i \omega_m ) }
\; \; \; .
\label{eq:frpadef}
\end{equation}
In the RPA screened interaction Eq.\ (\ref{eq:frpadef}) only the long wavelength
limit of $\Pi_0({\bf q}, i \omega_m)$ enters, consistent with the neglect
of the around-the-corner processes.
Thus, the ${\bf{q}}$ sums in Eqs.\ (\ref{eq:G0res})--(\ref{eq:Slondef})
are implicitly restricted to the regime
$| q_{\|}  | 
{ \raisebox{-0.5ex}{$\; \stackrel{<}{\sim} \;$}}
\lambda$,
$| q_{\bot}  | 
{ \raisebox{-0.5ex}{$\; \stackrel{<}{\sim} \;$}}
\Lambda$. However, as long as 
the external wave-vector ${\bf{q}}$ in Eq.\ (\ref{eq:Gbosres})
is sufficiently small, we may ignore these cutoffs.
Then it is easy to see that 
$ {{G}}^{\alpha}_{0} ( {\bf{r}}  , \tau  )$
in Eq.\ (\ref{eq:G0res}) is proportional to 
$\delta^{(d-1)} ( {\bf{r}}_{\bot} )$, where
the $d-1$-dimensional vector 
${\bf{r}}_{\bot}$  consists of the components of ${\bf{r}}$ that are
perpendicular to ${\bf{v}}^{\alpha}$.
In fact, for $V$ and $\beta \rightarrow \infty$ the integration in
Eq.\ (\ref{eq:G0res}) is easily performed analytically, with the result
 \begin{equation}
 G^{\alpha}_{0} ( {\bf{r}}  , \tau  )
 = 
 \delta^{(d-1)} ( {\bf{r}}_{\bot}  )
 \left( \frac{ - i}{2 \pi} \right)
 \frac{1}
 { 
 r_{\|}  
 + i | {\bf{v}}^{\alpha} |  \tau }
 \; \; \; ,
 \label{eq:Gpatchreal1}
 \end{equation}
where $r_{\|} = {\bf{r}} \cdot \hat{\bf{v}}^{\alpha}$.
Because of the  prefactor
 $\delta^{(d-1)} ( {\bf{r}}_{\bot}  )$, we may replace
${\bf{r}} \rightarrow r_{\|} \hat{\bf{v}}^{\alpha}$ 
in Eqs.\ (\ref{eq:Slondef}) and (\ref{eq:Qlondef}).
Furthermore the ${\bf{r}}$-integral in Eq.\ (\ref{eq:Gbosres}) is effectively
a one-dimensional one, because the integrations over the components
of ${\bf{r}}$ perpendicular to ${\bf{v}}^{\alpha}$
can be done trivially due to the
$\delta$-function in Eq.\ (\ref{eq:Gpatchreal1}).
Of course, this simplification is an artefact of the linearization.
The modifications of the above results due to
the non-linear terms in the energy dispersion 
have recently been calculated in Ref.\ \cite{Kopietz96}.
Most importantly, for non-linear energy dispersion 
one should replace in Eqs.\ (\ref{eq:Rlondef}) and (\ref{eq:Slondef})
 \begin{equation}
 \frac{1}{( i \omega_{m} - {\bf{v}}^{\alpha} \cdot {\bf{q}} )^{2 }}
 \rightarrow
 \frac{1}{( i \omega_{m} - \xi^{\alpha}_{\bf{q}} )
 ( i \omega_{m} + \xi^{\alpha}_{- \bf{q}} )}
 \label{eq:polereplace}
 \; \; \; ,
 \end{equation}
where $\xi^{\alpha}_{\bf{q}}$ is given in Eq.\ (\ref{eq:energy}).
Note that the non-linear terms 
remove the double pole in
Eqs.\ (\ref{eq:Rlondef}) and (\ref{eq:Slondef}). 
Moreover, non-linearities in the energy dispersion lead also to the
replacement of the 
function $G_0^{\alpha} ( {\bf{r}} , \tau )$ in Eq.\ (\ref{eq:Gbosres})
by an interaction-dependent Green's function 
$G_1^{\alpha} ( {\bf{r}} , \tau )$,
which does {\it{not}} involve
the singular
prefactor $\delta^{(d-1)} ( {\bf{r}}_{\bot} )$ 
given in Eq.\ (\ref{eq:Gpatchreal1}). 
For an explicit expression of
$G_1^{\alpha} ( {\bf{r}} , \tau )$ see
Refs.\ \cite{Kopietzhab} and \cite{Kopietz96}.

The above non-perturbative expression for the single-particle
Green's function are valid in arbitrary dimensions and for arbitrarily 
shaped Fermi surfaces. Furthermore, in $d=1$ these expressions 
correctly reduce to the well-known bosonization result of the
Tomonaga-Luttinger model.
We now apply the above results to study the problem of coupled
Luttinger liquids.

\section{ 
The Coulomb-interaction in chains  without interchain hopping}
\label{sec:thecoulomb}

Before discussing finite interchain hopping $t_{\bot}$, 
it is instructive  to study first the case $t_{\bot} = 0$, where the electrons 
are confined to the individual chains.
Of course, electrons on different chains still interact with the
three-dimensional Coulomb interaction, so that
this is not a purely one-dimensional problem. The latter
has been discussed in Ref.\ \cite{Schulz83}.
Thus, making the continuum approximation for motion parallel
to the chains, the Fourier transform of the bare interaction is given by
 \begin{equation}
 f_{\bf{q}} = e^2 a_{\bot}^2 \int_{- \infty}^{\infty} d r_x 
 {\sum_{\bf{r}_{\bot}}}^{\prime} 
 \frac{ e^{- i {\bf{q}} \cdot {\bf{r}} } }{  
 \sqrt{ r_x^2 + {\bf{r}}_{\bot}^2 } }
 \label{eq:fqbare}
 \; \; \; ,
 \end{equation}
where the ${\bf{r}}_{\bot}$-sum is over the  two dimensional lattice 
of chains, with interchain lattice spacing
$a_{\bot}$. The prime indicates
that the ${\bf{r}}_{\bot}= 0$ term must be properly 
regularized (see below).
For $ | {\bf{q}} | \ll a_{\bot}^{-1} $ Eq.\ (\ref{eq:fqbare})
reduces to the familiar result
$f_{\bf{q}} = 4 \pi e^2 / {\bf{q}}^2$.
Note that we do not replace by hand the long-range Coulomb interaction
by an effective screened short-range interaction. The screening problem
will be solved explicitly via our bosonization approach.

\begin{figure}[t]
\vspace{1.0cm}
\hspace{3.0cm}
\epsfysize6cm
\epsfbox{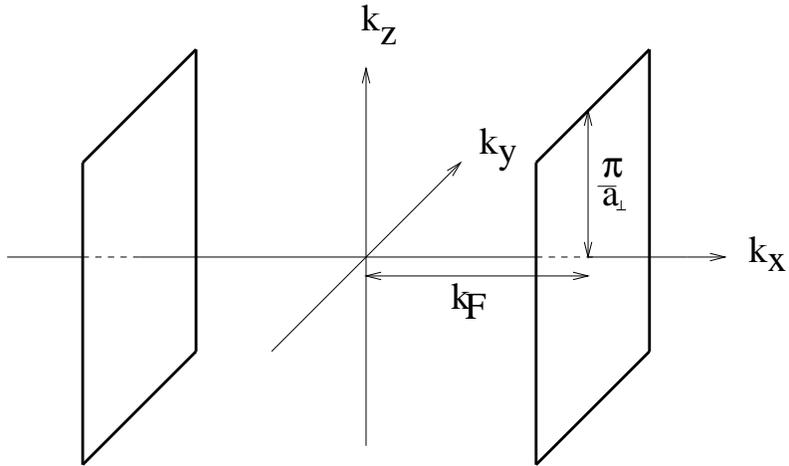}
\caption{ 
Fermi surface of an array of 
chains\index{Fermi surface!coupled chains ($t_{\bot} \neq 0$)}
without interchain hopping. Only the intersection with the plane $k_{z} = 0$ is shown. }
\vspace{1.0cm}
\label{fig:1d}
\end{figure}

Because for $t_{\bot} = 0$ the particle number is conserved on each chain,
the problem can be solved by means of
standard one-dimensional bosonization techniques \cite{Solyom,Schulz83,Penc93}. 
However, the solution can also be obtained quite elegantly
within the framework of higher-dimensional bosonization.
In the absence of interchain hopping, 
the Fermi surface consists of two parallel planes, as shown in Fig.\ \ref{fig:1d}. 
These planes can be identified with the patches discussed in 
Sec.\ \ref{sec:bos}.  Let us label the right plane
by $\alpha = +$, and the left one by $\alpha = -$.
The (linearized) energy dispersion is simply given by
$\xi^{\alpha}_{\bf{q}} = \alpha v_F q_x$, where
$v_F = | {\bf{v}}^{\alpha} |$ is the Fermi velocity. 
As usual, a dimensionless measure for the strength of the
interaction is $F_{\bf{q}} = \nu f_{\bf{q}}$, where
$\nu =  {2}/({ \pi v_{F} a_{\bot}^2 })$ is the
density of states at the Fermi energy (the factor of $2$ is due to the
spin-degeneracy). 
An important length scale in the problem is set by the
Thomas-Fermi screening wave-vector $\kappa$, which can be defined by
rewriting the  dimensionless interaction $F_{\bf{q}}$
in the long wavelength limit as $F_{\bf{q}} = \kappa^2 / {\bf{q}}^2$.
This yields
  $\kappa = a_{\bot}^{-1} \sqrt{8 \pi g }$,
where the dimensionless coupling constant $g$ is given by
 \begin{equation}
 g = \frac{e^2}{ \pi v_F}
 \label{eq:gdef}
 \; \; \; .
 \end{equation}
To evaluate Eqs.\ (\ref{eq:Qlondef})--(\ref{eq:Slondef}), we need to know
the RPA interaction in Eq.\ (\ref{eq:frpadef}),
which involves the non-interacting polarization
$\Pi_0 ( {\bf{q}} , i \omega_m )$.
In the absence of interchain hopping 
the polarization has the usual one-dimensional form,
which is in the long wavelength limit given by
 \begin{equation}
 \Pi_{0} ( {\bf{q}} , i \omega_m  ) = \frac{\nu}{2} \sum_{\alpha = \pm } \frac{ {\bf{v}}^{\alpha} \cdot {\bf{q}} }
 {  {\bf{v}}^{\alpha} \cdot {\bf{q}} - i \omega_{m} }
 = \frac{\nu }{ 1 + \omega_m^2 / (  v_F  q_x )^2 }
 \label{eq:Pi0patch}
 \; \; \; .
 \end{equation}
For better comparison with the calculations for finite interchain hopping
presented in Sec.\ \ref{sec:finiteinter},
it is convenient to express the RPA interaction in terms of the
dynamic structure factor in the usual way
(we take the limit $\beta \rightarrow \infty$)
 \begin{equation}
 f^{RPA} ( {\bf{q}} , i \omega_m ) 
  =  f_{\bf{q}} - f_{\bf{q}}^2 
 \int_{0}^{\infty} d \omega 
 S^{RPA} ( {\bf{q}} , \omega ) 
 \frac{  2 \omega}
 { \omega^{2} + \omega_{m}^2 }
 \label{eq:FrpaSRPA}
 \; \; \; ,
 \end{equation}
with
 \begin{equation}
 S^{RPA} ( {\bf{q}} , \omega ) = 
 \frac{1}{\pi} \mbox{Im} 
 \left\{
 \frac{\Pi_0 ( {\bf{q}} , \omega + i 0^{+} )}{ 1 +
 f_{\bf{q}} \Pi_0 ( {\bf{q}} , \omega + i 0^{+} )}
 \right\}
 \; \; \; .
 \label{eq:SPi}
 \end{equation}
Using Eq.\ (\ref{eq:Pi0patch}) we obtain
for $\omega > 0$
 \begin{equation}
 S^{RPA} ( {\bf{q}} , \omega ) =
 Z_{\bf{q}} \delta ( \omega - \omega_{\bf{q}} )
 \; \; \; ,
 \end{equation}
with the collective mode and the residue given by
\begin{eqnarray}
\omega_{{\bf{q}}} & = & \sqrt{ 1 + F_{\bf{q}} } v_{F} | q_{x} |
\; \; \; ,
\label{eq:col1patch}
\\
Z_{\bf{q}} & = & 
= \frac{ \nu v_{F} | q_{x} |}{2 \sqrt{ 1 + F_{\bf{q}} } }
=
\frac{ | q_{x} | }{\pi a_{\bot}^2 \sqrt{ 1 + F_{\bf{q}} } }
\label{eq:Zom1patch}
\; \; \; .
\end{eqnarray}
Substituting these results into Eq.\ (\ref{eq:Rlondef})
it is then easy to show that
 \begin{equation}
 R^{\alpha} =  -
 \frac{1}{V} \sum_{\bf{q}} f_{\bf{q}}^2 \frac{Z_{\bf{q}} }{ ( \omega_{\bf{q}} +
 | {\bf{v}}^{\alpha} \cdot {\bf{q}} | )^2 }
 \label{eq:Rres}
 \; \; \; ,
 \end{equation}
which for $V \rightarrow \infty$ reduces to
 \begin{equation}
 R^{\alpha} =  
 - \frac{1}{2} \int_{0}^{\infty}
 \frac{ d q_{x} }{q_{x}}
  \left< \frac{F_{\bf{q}}^2}
 {2 \sqrt{ 1 + F_{\bf{q}} } \left[ \sqrt{1+F_{\bf{q}} } + 1 \right]^2 }
 \right>_{BZ}
 \label{eq:R3patch}
 \; \; \; .
 \end{equation}
Similarly, we obtain
 \begin{eqnarray}
 {\rm Re} S^{\alpha} ( x , \tau )
 & = &  
 - \frac{1}{2} \int_{0}^{\infty}
 \frac{ d q_{x} }{q_{x}} 
 \cos ( q_{x} x )
 \left[ 
 \left<
 \frac{ 1  + \frac{F_{\bf{q}}}{2} }{  \sqrt{ 1 + F_{\bf{q}} } }
 e^{ - \sqrt{ 1 + F_{\bf{q}} } v_{F} q_{x} | \tau | } 
 \right>_{BZ}
 - e^{ - v_{F} q_{x} | \tau | } \right]
  \; ,
 \nonumber
 \\
 \label{eq:ReS3patch}
 \\
 {\rm Im} S^{\alpha} ( x , \tau )
 & = &
 - \frac{\mbox{sgn} ( \tau )}{2}
 \int_{0}^{\infty}
 \frac{ d q_{x} }{q_{x}} \sin ( q_{x}  x   )
 \left[ 
 \left< e^{ - \sqrt{ 1 + F_{\bf{q}} } v_{F} q_{x} | \tau | } \right>_{BZ}
 - e^{ - v_{F} q_{x} | \tau | } \right]
 \; \; ,
 \nonumber
 \\
 & &
 \label{eq:ImS3patch}
 \end{eqnarray}
where $x = \alpha r_x$, and
where for any function $h ( {\bf{q}} )$
the symbol $ \left< h ( {\bf{q}}  ) \right>_{BZ} $ denotes averaging over the
first transverse Brillouin zone (BZ), 
 \begin{equation}
 \left< h ( {\bf{q}} ) \right>_{BZ} = 
  \frac{1}{  ( \frac{2 \pi }{ a_{\bot}} )^2 }
 \int_{ - \frac{ \pi}{a_{\bot}} }^{ \frac{ \pi}{a_{\bot}} }   d q_{y} 
 \int_{ - \frac{ \pi}{a_{\bot}} }^{ \frac{ \pi}{a_{\bot}} }     d q_{z} 
 h ( {\bf{q}} )
 \label{eq:BZavdef}
 \; \; \; .
 \end{equation}
To show that  the $q_x$-integrals in Eqs.\ (\ref{eq:ReS3patch}) and 
(\ref{eq:ImS3patch}) exist at large $q_x$ without the addition of an ultraviolet
cutoff it is necessary to know the behavior of $F_{\bf q}$ for large $|{\bf q}|$. 
Trivially in the approximation $F_{\bf q}= \kappa^2/{\bf q}^2$ the interaction falls
off as $ |{\bf q}|^{-2}$.
If one takes into account that the one-dimensional continuous and 
two-dimensional discrete Fourier transform (\ref{eq:fqbare}) of the potential 
shows the same behavior (see below) it is easy to show from Eqs.\ (\ref{eq:ReS3patch}) 
and (\ref{eq:ImS3patch}) that after the BZ averaging the integrands for $\tau=0$ fall off 
like $(\kappa/|q_x|)^5$.
Because the $q_x$-integrals exist, $\kappa$ plays the role of an 
ultraviolet cutoff. For finite $\tau$ the 
integrands in Eqs.\ (\ref{eq:ReS3patch}) and (\ref{eq:ImS3patch}) only fall 
off like $(\kappa/q_x)^2$ but the $q_x$-integrals still exists.
 
Let us first consider the case
$\tau = 0$. Then we need to calculate the following Brillouin zone average
 \begin{equation}
 \gamma_{cb} ( {q_{x}} ) =
 \frac{1}{2}
 \left< \frac{F_{\bf{q}}^2}
 {2 \sqrt{ 1 + F_{\bf{q}} } [ \sqrt{1+F_{\bf{q}} } + 1 ]^2 }
 \right>_{BZ}
 =
 \frac{1}{2} \left[
  \left< \frac{ 1  + \frac{F_{\bf{q}}}{2} }{  \sqrt{ 1 + F_{\bf{q}} } } \right>_{BZ}
 - 1 \right]
 \label{eq:tildeFqparallel}
 \; \; \; .
 \end{equation}
In the regime $g \ll 1$ 
the Thomas-Fermi screening length $\kappa^{-1}$ is
large compared with the transverse lattice spacing $a_{\bot}$.
Because  in this case all wave-vector integrals are dominated by the regime
$| {\bf{q}} | 
{ \raisebox{-0.5ex}{$\; \stackrel{<}{\sim} \;$}} \kappa$
it is allowed to use 
the continuum approximation $f_{\bf q} = 4 \pi e^2 / {\bf{q}}^2$ 
for the Fourier transform of the Coulomb potential. 
The averaging over the transverse Brillouin zone in Eq.\ (\ref{eq:tildeFqparallel})
can then be done analytically, with the result
 \begin{equation}
 \gamma_{cb} ( {q_{x}} ) = 
 \frac{e^2}{2 \pi v_F}
 \frac{1}{ \left[
 | q_x | / \kappa + \sqrt{ 1 + ( q_x / \kappa )^2 } \right]^2 }
 \label{eq:gammacbqx}
 \; \; \; .
 \end{equation} 
Hence, 
 \begin{equation}
 Q^{\alpha} ( x , 0 ) =
 - \frac{e^2}{2 \pi v_F}
 \int_{0}^{\infty} d q_x 
 \frac{1 - \cos ( q_x x )}{ \left[
 | q_x | / \kappa + \sqrt{ 1 + ( q_x / \kappa )^2 } \right]^2 }
 \; \; \; .
 \end{equation}
For $x \gg \kappa^{-1}$ it is now easy to show that
 \begin{equation}
 Q^{\alpha} ( x , 0 ) \sim - \gamma_{cb} \ln [ \kappa x ]
 \label{eq:Qasym2patch}
 \; \; \; ,
 \end{equation}
with the anomalous dimension given by
 \begin{equation}
 \gamma_{cb} \equiv  
  \lim_{q_{x} \rightarrow 0}
 \gamma_{cb} ( q_{x} ) 
 =
  \frac{e^2}{ 2 \pi v_{F} }  
 \; \; \; .
 \label{eq:gammacbres}
 \end{equation}
The logarithmic growth of the static Debye-Waller factor
is one of the characteristics of a Luttinger liquid. 
Note that 
in a strictly one-dimensional model the long-range Coulomb interaction
leads to a Wigner crystal phase \cite{Schulz93} and not to a Luttinger
liquid. In this case the anomalous dimension diverges in the
thermodynamic limit. However, the Coulomb interaction  between electrons on different
chains removes this divergence, and leads to a Luttinger
liquid\cite{Schulz83} with modified spectral properties
for $k \neq k_F$ (see below).

We would like to point out that Eq.\ (\ref{eq:gammacbres}) is only valid 
for $g \ll 1$,  where
$\gamma_{cb} \ll 1$. It would be incorrect to extrapolate this
result to the regime where $g$ is of the order of unity,
which is  experimentally relevant. 
In order to calculate the anomalous dimension in this regime,
the Fourier transform $f_{\bf{q}}$ of the interaction has to be properly calculated.
For larger $g$ we have to take in to account the lattice structure 
in the transverse direction. To calculate the one-dimensional
continuous and two-dimensional discrete Fourier transform  
(\ref{eq:fqbare}) of
the Coulomb potential it is necessary to regularize the ${\bf
  r}_{\perp}=0$-contribution as in the strictly one-dimensional case
\cite{Schulz93}. The characteristic feature of the regularized ${\bf
  r}_{\perp}=0$-term is its logarithmic divergence $ - c \ln{ ( a_0 q_x 
)}$ for small $q_x$, where $a_0$ is the typical extension of
the one-particle wave-function in the transverse direction and $c$ is
a constant. The
prefactor and the behavior at larger momenta depend on the special
regularization procedure chosen. This divergence is
responsible for the divergence of the anomalous dimension in the
strictly one-dimensional model with Coulomb interaction, which leads to
the Wigner crystal phase \cite{Schulz93} of the one-dimensional electron
gas. For $g \ll 1$ we already saw that the chains coupled by the 
three-dimensional Coulomb interaction are {\it not} in the Wigner
crystal but in the Luttinger liquid phase. In the following we 
show that for arbitrary $g$ the term 
$ - c \ln{ ( a_0 q_x ) }$ 
for an infinite number of chains is exactly cancelled by a 
term $c \ln{ ( a_{\perp} q_x 
) }$ from the ${\bf r}_{\perp} \neq 0$ part of the Fourier
transform  (\ref{eq:fqbare}). Therefore the system is a 
Luttinger liquid for all $g$.

In our regularization procedure we assume that the one-particle wave
function in the transverse direction is given by a Gaussian
distribution corresponding to a harmonic confinement potential. We replace
the ${\bf r}_{\perp} = 0$-component in Eq.\ (\ref{eq:fqbare}) by 
\begin{eqnarray}
\label{potketten1}
&& \frac{a_{\perp}^2 e^2}{a_0^2 \pi} \int_{-\infty}^{\infty } 
d {\bf r}_{\perp} \int_{-\infty}^{\infty} dr_x  
\frac{e^{-iq_{x} r_x} e^{- {\bf r}_{\perp}^2/a_0^2}}{\left(r_x^2 +
{\bf r}_{\perp}^2\right)^{1/2}} \nonumber \\
&& = \frac{a_{\perp}^2 e^2}{a_0^2 \pi} \int_{-\infty}^{\infty } 
d {\bf r}_{\perp} K_0(|{\bf r}_{\perp}| q_x) e ^{- {\bf
    r}_{\perp}^2/a_0^2 } ,
\end{eqnarray}   
where $K_0(x)$ denotes the modified Bessel function. The logarithmic
divergence of $K_0$ for $x \to 0$ leads to the 
logarithmic behavior for small $q_x$ discussed above.

To obtain a fast convergence of the Fourier transform (\ref{eq:fqbare}) 
for ${\bf r}_{\perp} \neq 0$, we use the {\it Ewald summation
technique} \cite{Ziman}. With the help of the {\it Theta function
transformation}
\begin{eqnarray}
\label{thetafkttrafo}
\sum_{{\bf r}_{\perp}} e^{- \alpha |{\bf r}_{\perp}|^2 - i{\bf
 q}_{\perp} \cdot {\bf r}_{\perp}} 
= \frac{\pi}{a_{\perp}^2 \alpha^2} \sum_{{\bf G}_{\perp}}   
e^{-  |{\bf q}_{\perp} - {\bf G}_{\perp} |^2/(4 \alpha)} ,
\end{eqnarray}    
where ${\bf G}_{\perp}$ denotes a vector of the reciprocal lattice in
the transverse direction and  the integral 
\begin{eqnarray}
\label{einsdurchwurz}
\frac{2}{\sqrt{\pi}} \int_{0}^{\infty} d \alpha 
e^{- ( r_x^2 +|{\bf r}_{\perp}|^2)
 \alpha^2} =  \frac{1}{\sqrt{r_x^2+|{\bf r}_{\perp}|^2}} 
 \; \; \; ,
\end{eqnarray}
one obtains 
\begin{eqnarray}
\label{potketten3}
f_{\bf q} & = & 4 \pi  e^2 \sum_{{\bf G}_{\perp} }
\frac{ \exp{ \left\{ - \left[ q_{x}^2 + |{\bf q}_{\perp} - {\bf G}_{\perp} |^2
\right] /(4 \alpha_0) \right\} } }{ 
q_{x}^2 + |{\bf q}_{\perp} - {\bf G}_{\perp} |^2  }  \nonumber \\*
&& + 2 a_{\perp}^2 e^2 \int_{\alpha_0}^{\infty} \frac{d \alpha}{\alpha}
\sum_{{\bf r}_{\perp} \neq 0} e^{- \alpha |{\bf r}_{\perp}|^2
 - i{\bf q}_{\perp} \cdot {\bf r}_{\perp}}
\nonumber \\*
&& - 2 a_{\perp}^2 a_0^2 e^2 \int_{0}^{\alpha_0} d \alpha \alpha
\frac{1}{1+a_0^2 \alpha^2} e^{- q_{x}^2/(4 \alpha^2)} \nonumber \\*
&& + 2 a_{\perp}^2 e^2 \int_{\alpha_0}^{\infty} \frac{d \alpha}{\alpha}
\frac{1}{1+a_0^2 \alpha^2} e^{- q_{x}^2/(4 \alpha^2)} .
\end{eqnarray}                                                  
For $\alpha \in [0,\alpha_0]$ we used in  Eq.\ (\ref{einsdurchwurz}) 
the transformation Eq.\ (\ref{thetafkttrafo}) from the direct to the
reciprocal lattice. Now $\alpha_0$ can be chosen in such a way that in the
sums in the first and second line of Eq.\ (\ref{potketten3}) only a
few lattice vectors contribute. In the following we always
choose $\alpha_0 =4/a_0$. For this $\alpha_0$ the sum in the second
line in Eq.\ (\ref{potketten3}) can be completely neglected and in the
first line only the vectors ${\bf G}_{\perp} = (m_y 2 \pi/a_{\perp},
m_z 2 \pi/a_{\perp})$ with $m_y$ and $m_z \in [-5,5] $ contribute.
The last two terms in Eq.\ (\ref{potketten3}) contain the ${\bf
  r}_{\perp}=0$-term of Eq.\ (\ref{eq:fqbare}). 
From Eq.\ (\ref{potketten3}) it is easy to see that 
only for $|{\bf q} | \ll a_{\perp}^{-1}$   is is allowed to
approximate the Fourier transform of the Coulomb potential
by its continuum limit $ 4 \pi e^2/{\bf q}^2$.

\begin{figure}
\vspace{1cm}
\hspace{2cm}
\epsfysize11cm 
\epsfbox{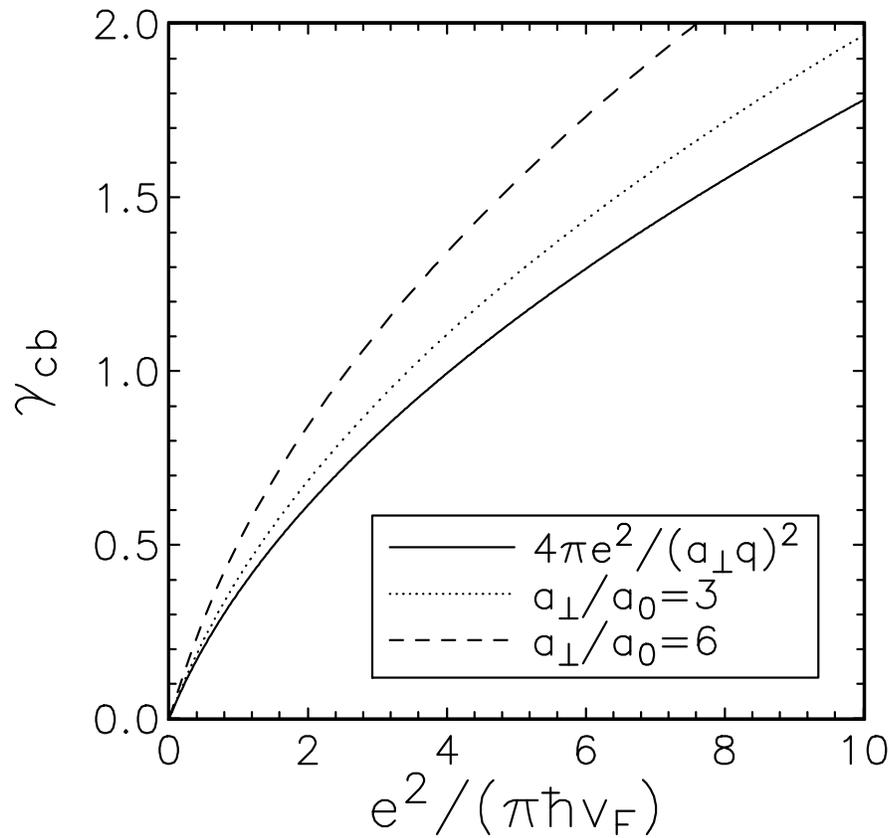}
\vspace{1cm}
\caption{The anomalous dimension $\gamma_{cb}$ as a function of the
coupling constant $g = e^2 / ( \pi \hbar v_F )$ for two values of $a_{\bot}/a_{0}$ (dashed and dotted
lines) and $f_{\bf q}=4 \pi e^{2}/(a_{\bot} {\bf{q}})^2$ 
(solid line).}
\label{fig:andim}
\end{figure}

To determine the anomalous dimension from Eqs.\ (\ref{eq:tildeFqparallel})
and (\ref{eq:gammacbres}),
we have to take the $q_x \to 0$ limit in Eq.\ (\ref{potketten3}).
For $q_x=0$ we obtain
\begin{eqnarray}
\label{potketten3qp0}
\tilde{v}(q_{x}=0,{\bf q}_{\perp}) & = & 4 \pi  e^2 \sum_{{\bf G}_{\perp} }
\frac{ \exp{ \left\{ -  |{\bf q}_{\perp} - {\bf G}_{\perp} |^2  
/(4 \alpha_0) \right\} } }{  |{\bf q}_{\perp} - {\bf G}_{\perp} |^2  }  \nonumber \\*
&& + 2 a_{\perp}^2 e^2 \int_{\alpha_0}^{\infty} \frac{d \alpha}{\alpha}
\sum_{{\bf r}_{\perp} \neq 0} e^{- \alpha |{\bf r}_{\perp}|^2 - 
i{\bf q}_{\perp} \cdot {\bf R}_{\perp}}
\nonumber \\*
&& - 2 a_{\perp}^2 e^2  \ln{\left[ 4 \frac{a_0}{a_{\perp}} \right] } ,
\end{eqnarray}       
i.e.\ no logarithmic divergence. As already discussed above, the
logarithmic divergence of the ${\bf r}_{\perp}=0$-term is exactly
cancelled. Therefore the anomalous dimension is finite and can be
calculated numerically following Eqs.\ (\ref{eq:tildeFqparallel}) and 
(\ref{eq:gammacbres}). In 
the regularization procedure we introduced the new parameter
$a_0$. Physically it is clear that only
ratios $a_{\perp}/a_{0}>1$ are relevant for
quasi-one-dimensional systems. 
In Fig.\ \ref{fig:andim} we present numerical results
for $\gamma_{cb}$  with $a_{\perp}/a_{0}=3$ (dotted line) and 
$a_{\perp}/a_{0}=6$ (dashed line). For comparison we also
show $\gamma_{cb}$ using the continuum approximation 
$f_{\bf q}=4 \pi e^2/{\bf q}^2$ (solid line).
From Eq.\ (\ref{potketten3}) it is easy to
show, that $F_{\bf q}$ falls off as
$\kappa^2/{\bf q}^2$ for large $|{\bf q}|$. Therefore our above
discussion concerning the  existence of the integrals
 Eqs.\ (\ref{eq:ReS3patch}) and 
(\ref{eq:ImS3patch}) is also applicable to the lattice Fourier transform.

In a Luttinger liquid the momentum integrated spectral function 
$\rho ( \omega )$, which
apart from a one-electron dipole matrix element determines angular 
integrated photoemission, is algebraically suppressed near the
chemical potential \cite{Luther74}
\begin{eqnarray}
\label{intspec}
\rho ( \omega ) \propto | \omega |^{{\gamma}_{cb}}.
\end{eqnarray}
Recent photoemission studies of quasi-one-dimensional conductors 
suggest values for the anomalous dimension in the range 
$1.0 \pm 0.2$ \cite{Dardel93,Nakamura94,Claessen95}. 
This would be hard to reconcile with a
model involving short range interactions, as e.g.\ the anomalous
dimension in the one-dimensional Hubbard model never exceeds $1/8$.
A rough estimate of the dimensionless coupling $g$ for experimental relevant
parameters following Ref.\ \cite{Jerome91} leads to 
values of $g$ of the order of one. 
As seen from Fig.\ \ref{fig:andim} our treatment using realistic Coulomb forces 
quite easily leads to anomalous dimensions in the experimental range. 

\begin{figure}
\epsfysize14cm 
\epsfbox{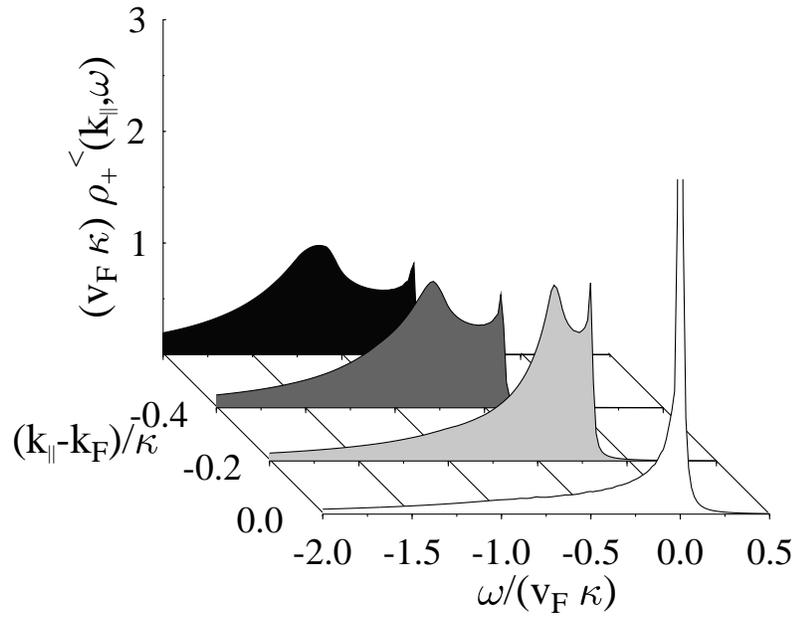}
\vspace{1cm}
\caption{The Lorentzian broadened spectral function $\rho_{+}^{<}$
for the coupled chains as a
function of $\omega$ for different $q_{\|}$. The parameters are
$g=0.8$, $a_{\bot}/a_{0}=3$ and the broadening $\chi /(v_{F}
\kappa)=0.01$. $\omega$ is measured relative to $\mu$ and $q_{\|}$
relative to $k_{F}$.}
\label{fig:specfu}
\end{figure}

In the following we discuss the momentum dependent 
spectral function $\rho^{<} ( q_{x} , \omega )$.
In the strictly
one-dimensional model with an interaction potential which is finite at
$q=0$ it is known that for anomalous dimensions $\gamma$
smaller than $1/2$ a power law divergence occurs at the energy $v_F
q$ and for $\gamma < 1$ at $v_c q$, where $v_c$ is the long wavelength
limit of the $q$-dependent charge excitation velocity (spin-charge separation)
\cite{Meden92,Voit93}. From Eq.\ (\ref{eq:col1patch}) one obtains for the 
coupled chains a ${\bf q}_{\perp}$-dependence of the charge velocity
\begin{eqnarray}
\label{cvelocity}
v_c({\bf q}_{\perp}) = \lim_{q_x \to 0} \frac{\omega_{\bf q}}{q_x} =
v_F \sqrt{1+F_{{\bf q}_{\perp}}}.  
\end{eqnarray}   
It is not obvious from Eqs.\ (\ref{eq:ReS3patch}) and (\ref{eq:ImS3patch}) 
whether a charge peak occurs or not, because 
the charge-velocity $v_{c} ( {\bf q }_{\perp} )$
is subject to the BZ averaging and the
singularity can be washed out.
On the other hand,
because the spin velocity $v_{F}$ is 
independent of ${\bf{q}}_{\bot}$,
one obtains for $\gamma_{cb} < \frac{1}{2}$ 
a sharp threshold singularity at $\omega = - v_{F} | q_{x}|$,
just like in one dimension,
\begin{eqnarray}
\label{rhokkettenvfs}
\rho_+^<(q_{x},\omega) \sim \Theta(v_F [q_{x}-k_F] -\omega) 
(v_F [q_{x}-k_F] -\omega)^{\gamma_{cb}-1/2} .
\end{eqnarray}     
For $\gamma_{cb} > \frac{1}{2}$ the singularity
is washed out, but the threshold survives. On the Fermi surface the BZ
averaging is irrelevant and one obtains for $\gamma_{cb} <1$
\begin{eqnarray}
\label{rhokfketten}
\rho_+^<(k_F,\omega) \sim \Theta(-\omega) (-\omega)^{\gamma_{cb}-1} 
\end{eqnarray}
as in the one-dimensional Tomonaga-Luttinger model.
In order to evaluate the momentum dependent spectral function
for arbitrary $q_x$ and $\omega$, we have to use 
numerical techniques developed in Ref.\ \cite{Meden93}. 
It turns out that a {\it broadened} charge peak occurs. For increasing
anomalous dimensions, i.e.\ increasing $g$, the charge peak
broadens. To clearly demonstrate spin-charge separation, we 
present in Fig.\ 8 spectra for $\gamma_{cb} =
0.33$, corresponding to $g=0.8$, with 
$a_{\perp}/a_0=3$. The spectra are calculated for large but finite
systems. The system size is given by the dimensionless level spacing
$\nu_{\kappa}= (2\pi/L)/\kappa$. In Fig.\ \ref{fig:specfu} we chose 
$\nu_{\kappa} = 1/80$. 
For technical reasons we multiplied the Green's function by a
factor $\exp{ \left\{ - \chi |t| \right\} }$ before we performed the
time-frequency Fourier transformation. This leads to a Lorentzian
broadening of the discrete spectral function and smears out the
threshold behavior at $\omega=  v_F (k_x-k_F)$. As can be seen from
Fig.\ \ref{fig:specfu}, the artificial broadening $\chi/(v_F \kappa) = 0.01$ is much
smaller than the natural broadening of the charge peak. The power law
behavior of the spectra at $\omega=  v_F (k_x-k_F)$
leads to an asymmetry of the related peaks in Fig.\ \ref{fig:specfu}. 
We checked that for a given $\chi$ the numerical spectra do not
change if the system size is further increased. In the numerical
calculations we have to cutoff the $q_x$-integrals in Eqs.\ 
(\ref{eq:ReS3patch}) and (\ref{eq:ImS3patch})
at $q_x=  n \kappa$, with some number $n$. In Fig.\ 8 we chose $n=5$ but we checked
that for the momenta and energies presented the
spectra are the same for larger values of $n$.  
We found that the charge peak for $-0.5 < (k_x-k_F)/
\kappa < 0$ disperses like $\omega = \bar{v}_c (k_x-k_F) $ with an
effective $g$ dependent charge velocity $\bar{v}_c$ which has to be determined 
numerically. For anomalous dimensions larger than one no
divergence at $\omega=  v_F (k_x-k_F)$ occurs and the spectra are
dominated by a very broad charge peak.   

To summarize this section, we showed that the system of one-dimensional
chains coupled by the long ranged Coulomb interaction is a Luttinger liquid
with an anomalous dimensions of the order of one. Furthermore the charge 
singularity of the strictly one-dimensional Luttinger liquid is broadened due to
the BZ averaging, but is still present. In the following we extend the model
and include transverse interchain hopping.

\section{Finite interchain hopping}
\label{sec:finiteinter}

\begin{figure}[t]
\vspace{1.0cm}
\hspace{3.0cm}
\epsfysize6cm
\epsfbox{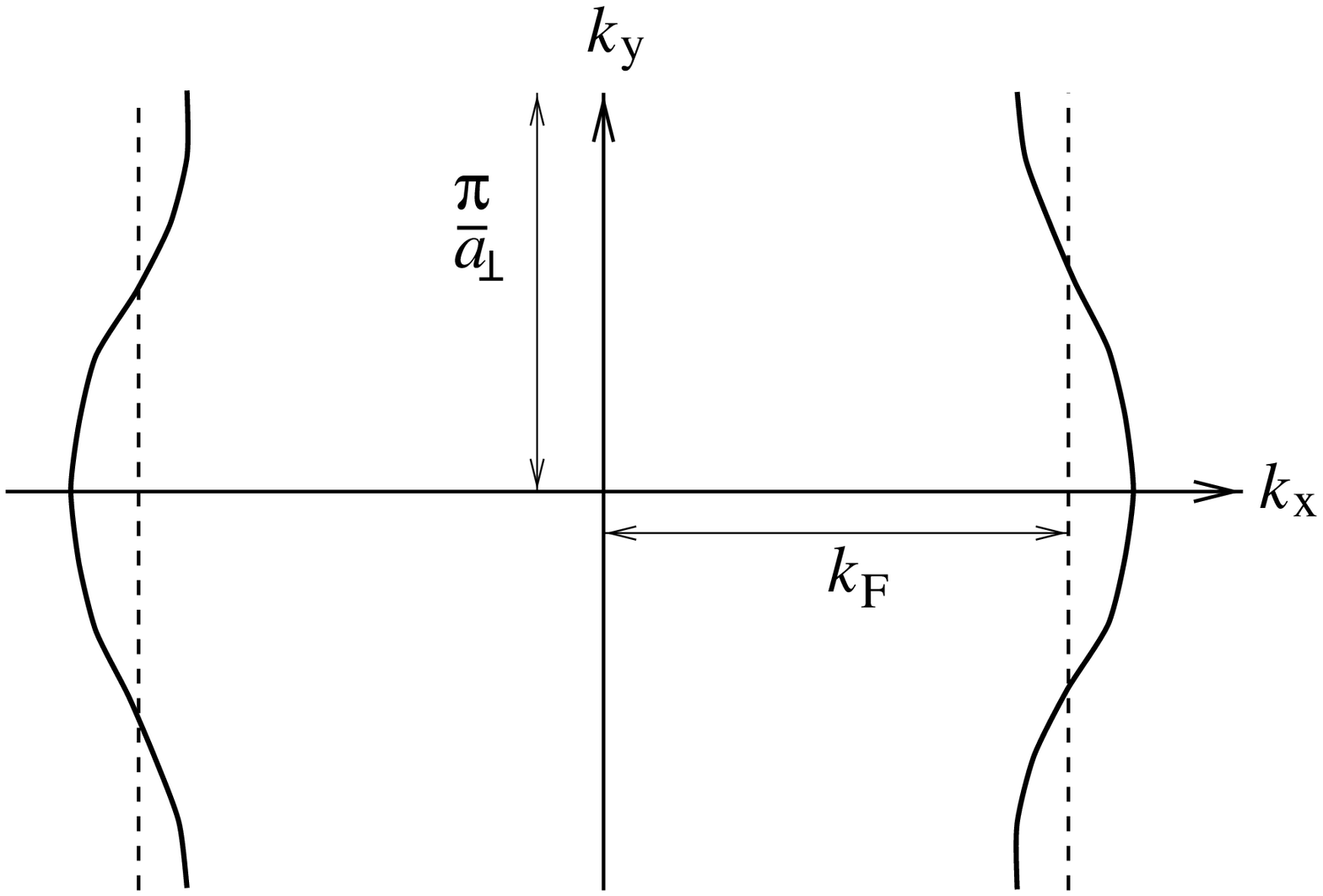}
\caption{ 
Fermi surface of an array of 
chains\index{Fermi surface!coupled chains ($t_{\bot} \neq 0$)}
with small interchain hopping. Only the intersection with the plane $k_{z} = 0$ is shown. }
\vspace{1.0cm}
\label{fig:quasi1d}
\end{figure}

\begin{figure}[h]
\vspace{1.0cm}
\hspace{3.0cm}
\epsfysize6cm
\epsfbox{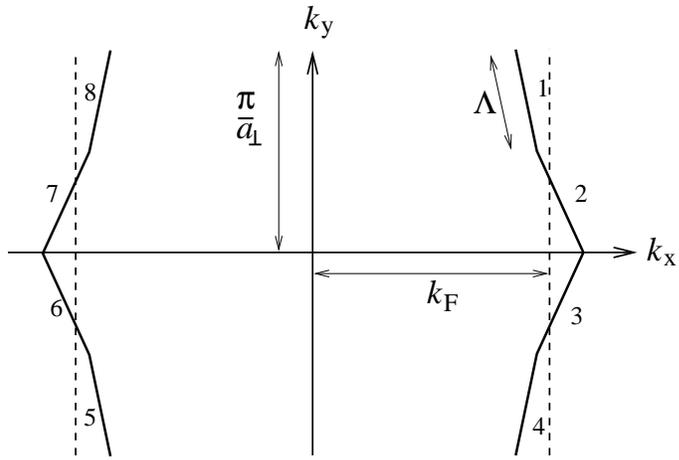}
\caption{Approximation of the Fermi surface in Fig.\ref{fig:quasi1d}
by $M = 8 $ flat patches.}
\vspace{1.0cm}
\label{fig:quasi1dpatch}
\end{figure}

\begin{figure}[h]
\vspace{1.0cm}
\hspace{3.0cm}
\epsfysize6cm
\epsfbox{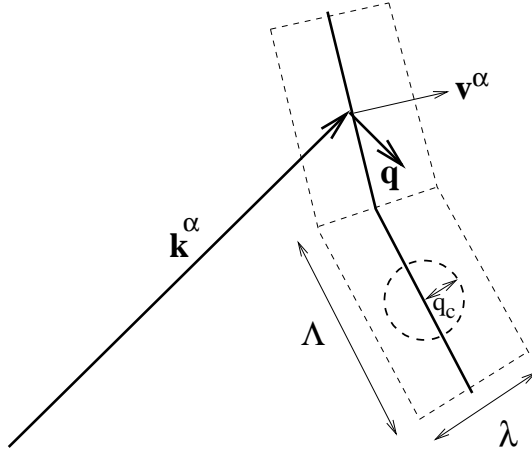}
\caption{ 
Squat boxes associated with the first two patches in Fig.\ref{fig:quasi1dpatch}.
The dashed circle indicates the maximal possible momentum
transfer $q_c$ of the interaction. The vector ${\bf{k}}^{\alpha}$ points
to the origin of a local coordinate system centered at patch $P^{\alpha}_{\Lambda}$, and
${\bf{v}}^{\alpha}$ is the corresponding local Fermi velocity.
}
\vspace{1.0cm}
\label{fig:squat}
\end{figure}

A typical  Fermi surface  of an array of chains with
small finite interchain hopping is shown in Fig.\ \ref{fig:quasi1d}.
In the quasi-one-dimensional materials discussed in Ref.\ \cite{Jerome91}
the intrachain hopping $t_x$ is an order of magnitude larger than
the interchain hopping $t_y = t_{\bot}$, which is again a factor of $10$ larger
than $t_z$. We therefore assume transverse hopping only in the $y$ direction.
In this section we shall apply our 
higher-dimensional bosonization approach to this problem.
We would like to emphasize that we are ultimately interested in 
realistic Fermi surfaces without nesting symmetries.
Unfortunately, by replacing 
a curved Fermi surface of the type shown in
Fig.\ \ref{fig:quasi1d} by 
a finite number of locally 
flat patches (see Fig.\ \ref{fig:quasi1dpatch}), we introduce unphysical
nesting symmetries. 
In order to obtain physical results that can be compared to
experiments, the singularities caused by these 
artificial symmetries should therefore be separated 
from the physical plasmon mode in the dynamic structure factor.
In this section we shall show how this 
can be done in practice.

\subsection{The $M$-patch model}
\label{subsec:Mpatch}

In the higher-dimensional bosonization approach with linearized energy dispersion
the Fermi surface is approximated by a finite number $M$ of flat
patches $P^{\alpha}_{\Lambda}$. 
In order not to break the inversion symmetry of the Fermi surface,
we choose an even number of patches of equal size
such that for each $P^{\alpha}_{\Lambda}$ there exists
another patch $P^{\bar{\alpha} }_{\Lambda}$ such that
the local Fermi velocities satisfy ${\bf{v}}^{\bar{\alpha}} = - {\bf{v}}^{\alpha}$
(see Fig.\ \ref{fig:quasi1dpatch} for $M = 8$). 
Then 
the non-interacting polarization
$\Pi_{0} ( {\bf{q}} , z )$ is at long wavelengths
given by 
 \begin{equation}
 \Pi_{0} ( {\bf{q}} , z )  =  \frac{ 2 \nu}{M} \sum_{\alpha = 1}^{M/2}
 \frac{ ( {\bf{v}}^{\alpha} \cdot {\bf{q}} )^2 }{
 ( {\bf{v}}^{\alpha} \cdot {\bf{q}})^2 - z^2 }
 \label{eq:Pi0Meven}
 \; \; \; ,
 \end{equation}
where it is understood that the sums are over all patches with
${\bf{v}}^{\alpha} \cdot {\bf{q}} \geq 0$.
For finite $M$ the poles of the RPA interaction, 
which are simply the zeros of
 \begin{equation}
 1 + f_{\bf{q}}
 \Pi_0 ( {\bf{q}} , z ) = 0
 \; \; \; ,
 \label{eq:PolynomM}
 \end{equation}
can be easily obtained by plotting
the right-hand side of Eq.\ (\ref{eq:Pi0Meven}) as function of $\omega^2$
where $\omega=z $ is real, and looking for the intersections with $-1 / f_{\bf{q}}$.
For generic ${\bf{q}}$ all
$ ( {\bf{v}}^{\alpha} \cdot {\bf{q}}  )^2$ are different and positive, and we can
order
$ 0 <
 ( {\bf{v}}^{\alpha_1} \cdot {\bf{q}} )^2 <
 ( {\bf{v}}^{\alpha_2} \cdot {\bf{q}} )^2 < \ldots
 <
 ( {\bf{v}}^{\alpha_{M/2}} \cdot {\bf{q}} )^2$. 
A repulsive interaction $f_{\bf{q}}$ leads then to 
zeros
$(\omega_{\bf{q}}^2)^{ ( \alpha )}$, $\alpha = 1 , \ldots , M/2$ of Eq.\ 
(\ref{eq:PolynomM}) as function of $z^2$ lying between the unperturbed poles,
 \begin{equation}
 ( {\bf{v}}^{\alpha_1} \cdot {\bf{q}} )^2 <
 ( \omega_{\bf{q}}^2 )^{(1)} <
 ( {\bf{v}}^{\alpha_2} \cdot {\bf{q}} )^2 < \ldots
 <
 ( {\bf{v}}^{\alpha_{M/2}} \cdot {\bf{q}} )^2 < 
 ( \omega_{\bf{q}}^2 )^{(M/2)} 
 \; \; \; .
 \end{equation}
In the large  $M$-limit the first $M/2 - 1$ solutions form the
particle-hole quasi-continuum, while the largest mode
$\omega_{\bf{q}}^{M/2}$ can be identified with the collective
plasmon mode.
For ${\bf{q}}$-values which lead to ${\bf{v}}^{\alpha_i} \cdot {\bf{q}} =0$
and (or)
$( {\bf{v}}^{\alpha_i} \cdot {\bf{q}} )^2 = 
( {\bf{v}}^{\alpha_j} \cdot {\bf{q}} )^2 $ the number
of poles in the quasi-continuum with non-vanishing residue is reduced.
Denoting by $Z_{\bf{q}}$ the residue of the plasmon mode,
the dynamic structure factor is given by
 \begin{equation}
 S^{RPA} ( {\bf{q}} , \omega ) = Z_{\bf{q}} \delta ( \omega - \omega_{\bf{q}} )
 + S^{RPA}_{incoh} ( {\bf{q}}, \omega )
 \label{eq:Srpaincoh}
 \; \; \; ,
 \end{equation}
where
  $S^{RPA}_{incoh} ( {\bf{q}}, \omega )$ is due to the
contribution of the other modes, which merge
for $M \rightarrow \infty$ into the
particle-hole continuum.
In the absence of interchain hopping 
  $S^{RPA}_{incoh} ( {\bf{q}}, \omega ) = 0$. Furthermore, 
the plasmon mode and the associated residue 
are given in
Eqs.\ (\ref{eq:col1patch}) and (\ref{eq:Zom1patch}).

\subsubsection{The plasmon mode}

The crucial observation is now that for small but finite
$t_{\bot}$ and $M \to \infty$ the low-energy 
behavior of the Green's function is
still dominated by the plasmon mode, which therefore describes the crossover
between Luttinger and Fermi liquid behavior.
Thus, to leading order in $\theta$, we may 
ignore the contribution 
  $S^{RPA}_{incoh} ( {\bf{q}}, \omega )$ in Eq.\ (\ref{eq:Srpaincoh}).
In this case the constant part $R^{\alpha}$ of the Debye-Waller factor
is still given by Eq.\ (\ref{eq:Rres}),
but now with $\omega_{\bf{q}}$ and $Z_{\bf{q}}$ given by the
plasmon mode of the $M$-patch model.
In the strong coupling limit it is easy to obtain an analytic
expression for $\omega_{\bf{q}}$ and $Z_{\bf{q}}$.
Anticipating that at strong coupling there exists
a pole $\omega_{\bf{q}}$  of Eq.\ (\ref{eq:PolynomM})
with $\omega_{\bf{q}}^2 = O ( F_{\bf{q}} )$, we may 
expand $\Pi_{0} ( {\bf{q}} , z )$ in powers of $z^{-2}$.
The leading term is
 \begin{eqnarray}
\Pi_{0} ( {\bf{q}} , z ) & = &
-  
\frac{ 2 \nu }{M} \sum_{\alpha = 1}^{M/2}
 \left( \frac{ {\bf{v}}^{\alpha} \cdot {\bf{q}} }{z} \right)^2 
  \left[ 1 +  \left( \frac{ {\bf{v}}^{\alpha} \cdot {\bf{q}} }{z} \right)^2 
+ O \left( \left[ 
  \frac{ {\bf{v}}^{\alpha} \cdot {\bf{q}} }{z} \right]^4   \right) 
\right]  
 \label{eq:Pi0zlargeexp}
 \; \; \; .
 \end{eqnarray}
Substituting this approximation into Eq.\ (\ref{eq:PolynomM}),
it is easy to show that 
 \begin{eqnarray}
 \omega_{\bf{q}}^2 & = &
 F_{\bf{q}} 
\frac{ 2}{M} \sum_{\alpha = 1}^{M/2}
 ( {\bf{v}}^{\alpha} \cdot {\bf{q}} )^2
 + \frac{ \sum_{\alpha = 1}^{M/2}
 ( {\bf{v}}^{\alpha} \cdot {\bf{q}} )^4 }{
 \sum_{\alpha = 1}^{M/2}
 ( {\bf{v}}^{\alpha} \cdot {\bf{q}} )^2} + O \left( 1/F_{\bf q} \right)
 \label{eq:plasmonpatchlarge}
 \end{eqnarray}
and
 \begin{equation}
 Z_{\bf{q}} 
 =  \frac{\nu}{2 \sqrt{F}_{\bf{q}} }
 \left[ 
\frac{ 2}{M} \sum_{\alpha = 1}^{M/2}
 ( {\bf{v}}^{\alpha} \cdot {\bf{q}} )^2  \right]^{1/2} 
+ O \left( 1/F_{\bf q} \right)
 \label{eq:Zqclose}
 \; \; \; .
 \end{equation}
Introducing these expressions in Eq.\ (\ref{eq:Rres}) we obtain to leading 
order in the coupling
\begin{eqnarray}
\label{eq:srpapl}
S_{pl}^{RPA}({\bf q},\omega) \approx \frac{\nu v_F}{2 \sqrt{F_{\bf q}}} 
\sqrt{q_x^2 + \eta^2 {\bf q}_{\perp}^2} 
\delta \left( \omega -
v_F \sqrt{F_{\bf q}} 
\sqrt{q_x^2 + \eta^2 {\bf q}_{\perp}^2}
\right) \;\; ,
\label{eq:Srpapl}
\end{eqnarray}
where 
\begin{eqnarray}
\label{eq:etadef}
\eta^2 = \frac{2}{M} \sum_{\alpha =1}^{M/2} \left( \hat{\bf v}^{\alpha} \cdot 
\hat{\bf q}_{\perp} \right)^2
\end{eqnarray}
is a small parameter for a weakly corrugated Fermi surface corresponding
to small interchain hopping.  Note that 
${\bf{q}}_{\bot}$ in Eq.(\ref{eq:Srpapl})
denotes the $q_y$- and $q_z$-components of ${\bf{q}}$, and should
not be confused with the vector ${\bf{q}}^{\alpha}_{\bot}$
in Eq.(\ref{eq:energy}).
The contribution to the constant part 
$R_{pl}^{\alpha}$ of the
Debye-Waller factor is given by 
\begin{eqnarray}
\label{eq:rplasmon} 
 R^{\alpha}_{pl} =  - \frac{1}{2} \int_{0}^{ \infty } d q_{x}
 \left< 
 \frac{1}{ \sqrt{ q_{x}^2 + \eta^2 q_{y}^2 }} 
  \frac{ \sqrt{F_{\bf{q}}} }{ \left[ 1+ \frac{1}{\sqrt{F_{\bf q}}} 
\frac{ {\bf{v}}^{\alpha} \cdot {\bf q}}{v_F \sqrt{q_x^2 + \eta^2 
{\bf q}_{\perp}^2}} 
 \right]^2 }
 \right>_{BZ}
 \; \; \; .
 \end{eqnarray}
To leading order the patch dependence drops out. For $\eta=0$, corresponding
to zero interchain hopping the integral is logarithmically divergent. For 
finite $\eta$ the $q_x$ integration obtains a finite lower cutoff which
leads to a contribution $\ln{[ \kappa / (\eta |{\bf q}_{\perp}|) 
] }$. To leading order in $\ln{(1/\eta)}$ one can take the limit
$q_x \rightarrow 0$ in $F_{\bf q}$. This yields 
\begin{eqnarray}
\label{eq:rplasmonapp}
 R^{\alpha}_{pl} \approx  \frac{1}{2}
\ln{(1/\eta)} \lim_{q_x \to 0} \left< \sqrt{F_{\bf q}} \right>_{BZ}
 \; \; \; .
 \end{eqnarray}
The prefactor of the logarithm is just the strong coupling expression 
of the anomalous dimension of the zero hopping $\eta=0$ limit.
Now, if $R_{pl}^{\alpha}$ is finite, the Riemann-Lebesgue Lemma 
\cite{Bender} guarantees
that $S_{pl}^{\alpha}(r_{\parallel} \hat{{\bf v}}_{\alpha},0)$ goes
to zero for increasing $r_{\parallel}$. Therefore the system is no longer
a Luttinger liquid, but a Fermi liquid with a finite quasi-particle weight
\cite{Kopietz94,KMS95}
\begin{eqnarray}
\label{quasiweightdef}
Z^{\alpha} = e^{R^{\alpha}} \;\; .
\end{eqnarray} 

\subsubsection{The nesting mode}

Clearly,
by approximating the realistic curved Fermi surface 
shown in Fig.\ \ref{fig:quasi1d} by 
a collection of  flat patches, 
we have introduced an artificial nesting symmetry.
For example, the patches $P^{\alpha}_{\Lambda}$ and 
$ P^{\alpha + 4}_{\Lambda}$ 
(with $\alpha = 1, 2,3,4$)
in Fig.\ \ref{fig:quasi1dpatch} 
can be connected by constant vectors 
which can be attached to an arbitrary point on the patches.
Simple perturbation theory \cite{Hlubina94}
indicates that this nesting symmetry
gives rise to logarithmic singularities,
leading to a breakdown of the Fermi liquid state.
However, unless there exists a real physical
nesting symmetry in the problem, 
this singularity has been
artificially generated by approximating  
a curved Fermi surface by flat patches.

Let us first show how these nesting singularities 
manifest themselves within our bosonization approach, and then 
give a simple quantitative argument how these singularities disappear
due to curvature effects.
The crucial observation is that
for a given ${\bf{q}}$
there exists one  special patch $P^{\beta}_{\lambda}$ such that
the energy $ | {\bf{v}}^{{\beta}} \cdot {\bf{q}} |$ is much
smaller than all the other energies
$ | {\bf{v}}^{\alpha} \cdot {\bf{q}} |$,
$\alpha \neq {\beta}$. 
Anticipating that for sufficiently small
$q_{\|} \equiv \hat{\bf{v}}^{{\beta}} \cdot {\bf{q}}$ 
there exists a $\delta$-function peak in the dynamic structure factor
at the nesting mode $\omega_{\bf{q}}^{{\beta}} \propto |{\bf{v}}^{ {\beta}} 
\cdot {\bf{q}}|$, we see that
for small $q_{\|}$ 
the energy $\omega_{\bf{q}}^{{\beta}} $ is much smaller than
all  energies $ | {\bf{v}}^{\alpha} \cdot {\bf{q}} |$
with $\alpha \neq {\beta}$. 
Hence, the energy dispersion of the nesting  mode can be
approximately calculated by setting $z^2=0$ in all terms 
with $\alpha \neq {\beta}$ in the sum Eq.\ (\ref{eq:Pi0Meven}).
This yields in the regime
of wave-vectors defined by
 \begin{equation}
 | {\bf{v}}^{{\beta}} \cdot {\bf{q}} | \ll
 | {\bf{v}}^{\alpha} \cdot {\bf{q}} |
 \; \; \; 
 \mbox{for all
 $ \alpha \neq {\beta}$}
 \label{eq:regime}
 \end{equation}
the following expression for the polarization
 \begin{equation}
 \Pi_{0} ( {\bf{q}} , z ) \approx
 \frac{\nu}{M}
 \left[ M-2 + \frac{ 2 ( {\bf{v}}^{{\beta}} \cdot {\bf{q}} )^2}{
  ( {\bf{v}}^{ {\beta} } \cdot {\bf{q}} )^2 - z^2}
  \right]
  \label{eq:nestpolarization}
  \; \; \; .
  \end{equation}
The collective mode Eq.\  (\ref{eq:PolynomM}) is
then easily solved, with the result that the dispersion of the
nesting mode is given by
 \begin{equation}
 \omega_{\bf{q}}^{ {\beta}} = 
 \sqrt{ \frac{1 + F_{\bf{q}} }{1 +
 \frac{M-2}{M} F_{\bf{q}} }} | {\bf{v}}^{{\beta}} \cdot {\bf{q}} |
 \label{eq:nestM}
 \; \; \; .
 \end{equation}
For the associated residue
we obtain
 \begin{eqnarray}
 Z_{\bf{q}}^{{\beta}} & = & 
  \frac{ \nu | {\bf{v}}^{{\beta}} \cdot {\bf{q}} |}{
 M \left[ 1 + \frac{M-2}{M} F_{\bf{q}} \right]^{\frac{3}{2}}
 \left[ 1 + F_{\bf{q}} \right]^{\frac{1}{2}} }
 \label{eq:nestres}
 \; \; \; .
 \end{eqnarray}
Using Eq.\ (\ref{eq:Rres}), we see that the nesting mode 
gives rise to the following contribution to the
constant part $R^{\beta}$ of the Debye-Waller factor associated with patch
$P^{\beta}_{\Lambda}$,
 \begin{equation}
 R^{\beta}_{nest} = 
 \frac{1}{V \nu M  } {\sum_{\bf{q}}}^{\prime} 
  \frac{ F_{\bf{q}}^2 }{
  | {\bf{v}}^{\beta} \cdot {\bf{q}} | 
 \left[ 1 + \frac{M-2}{M} F_{\bf{q}} \right]^{\frac{3}{2}}
 \left[ 1 + F_{\bf{q}} \right]^{\frac{1}{2}} 
 \left[
 \sqrt{ \frac{1 + F_{\bf{q}} }{1 +
 \frac{M-2}{M} F_{\bf{q}} }} + 1 \right]^2 }
 \label{eq:Rnest}
 \; \; \; ,
 \end{equation}
where the prime on the sum indicates that the
sum is over the wave-vector regime defined in Eq.\ (\ref{eq:regime}).
To understand the geometric meaning of Eq.\ (\ref{eq:regime}),
let us note that
for our $M$-patch model the angle between ${\bf{v}}^{\beta}$ and
the neighboring velocity ${\bf{v}}^{\beta + 1}$ is 
of the order of $\theta / M$, where
 \begin{equation}
 \theta = \frac{| {t}_{\bot} |}{E_{F}}
 \; \; \; .
 \label{eq:Thetasmallpardef}
 \end{equation}
Setting $q_{\|} = \hat{\bf{v}}^{\beta} \cdot {\bf{q}}$ and
$q_{\bot} = {\hat{\bf{v}}}^{\beta}_{\bot} \cdot {\bf{q}}$
(where ${\hat{\bf{v}}}^{\beta}_{\bot} $ is a unit vector orthogonal to
$ {\hat{\bf{v}}}^{\beta} $) it is easy to see that
the condition Eq.\ (\ref{eq:regime})
is equivalent with $| q_{\|} | \ll \theta | q_{\bot} | / M$.
We conclude that
for $V \rightarrow \infty$
the integral in Eq.\ (\ref{eq:Rnest}) is 
proportional to
 \begin{equation}
 R^{\beta}_{nest}  \propto
 \frac{1}{M} 
 \int_{0}^{\kappa} d q_{\bot}
 \int_{0}^{\theta  q_{\bot} / M} \frac{d q_{\|} }{ q_{\|}}
\label{eq:nestint}
\; \; \; ,
\end{equation}
which is infrared divergent.
Obviously
the logarithmic divergence is removed
if we take the limit $M \rightarrow \infty$.
It should be kept in mind, however, that
in the limit $M \rightarrow \infty$ 
the size of the patches vanishes, so that
the condition $q_{c} \ll \mbox{min} \{ \Lambda , \lambda \}$
is violated.
As discussed in Sec.\ \ref{sec:bos}, this condition is
necessary to justify the neglect of
the around-the-corner processes in the higher-dimensional bosonization 
approach, see Fig.\ \ref{fig:squat}.
Nevertheless, because for vanishing patch cutoff we 
recover a curved Fermi surface, the above calculation suggests that
the nesting singularity will disappear as soon as the finite
curvature of the patches is taken into account.
In Sec.\ \ref{subsubsec:nesting} we shall
use the higher-dimensional bosonization result for curved
patches \cite{Kopietzhab,Kopietz96} to show that this is indeed the case.
Here we  would like to give a simple quantitative argument which
captures the essential physics.
Suppose we retain the quadratic terms in the
expansion Eq.\ (\ref{eq:energy}) of the energy dispersion close to 
patch $P^{\beta}_{\Lambda}$.
Obviously the term
$q_{\bot}^2 / ( 2 m_{\bot} )$ (describing
the curvature of patch $P^{\beta}_{\Lambda}$) becomes
important for
$v_F  | q_{\|} | \approx 
q_{\bot}^2 / ( 2 m_{\bot} )$,
where
for simplicity we have assumed that $m_{\bot}^{\beta} = m_{\bot}$ and 
$| {\bf{v}}^{\beta} | = v_F$ are independent of the patch index $\beta$.
Hence,
we expect that for curved patches 
the lower limit for the $q_{\|}$-integral 
in Eq.\ (\ref{eq:nestint})
will be effectively replaced by
$\frac{q_{\bot}^2}{2 | m_{\bot} | v_F  }$.
We conclude that the effect of curvature 
can be qualitatively taken into account by substituting
 \begin{equation}
\int_{0}^{ \theta q_{\bot} / M } \frac{d q_{\|}}{  q_{\|}  }
\rightarrow
\int_{
\frac{q_{\bot}^2}{2 | m_{\bot} | v_F   }
}^{ \theta q_{\bot} /M } \frac{d q_{\|}}{  q_{\|}  }
= \ln \left[ \frac{ 2 | m_{\bot} | v_F  
\theta}{  q_{\bot}  M } \right]
\label{eq:curvaturesub}
\; \; \; .
\end{equation}
In physically relevant cases we expect 
$| m_{\bot} | \approx  k_F / ( v_F  \theta )$, 
so that the right-hand side of Eq.\ (\ref{eq:curvaturesub}) reduces to the
integrable factor of $\ln [ 2 k_F / ( M | q_{\bot}|) ]$.
A more rigorous justification for the regularization
given in Eq.\ (\ref{eq:curvaturesub}) will be given
in Sec.\ \ref{subsubsec:nesting}.

\subsection{The $4$-patch model}

\begin{figure}[hit]
\vspace{1.0cm}
\hspace{1cm}
\epsfysize6cm
\epsfbox{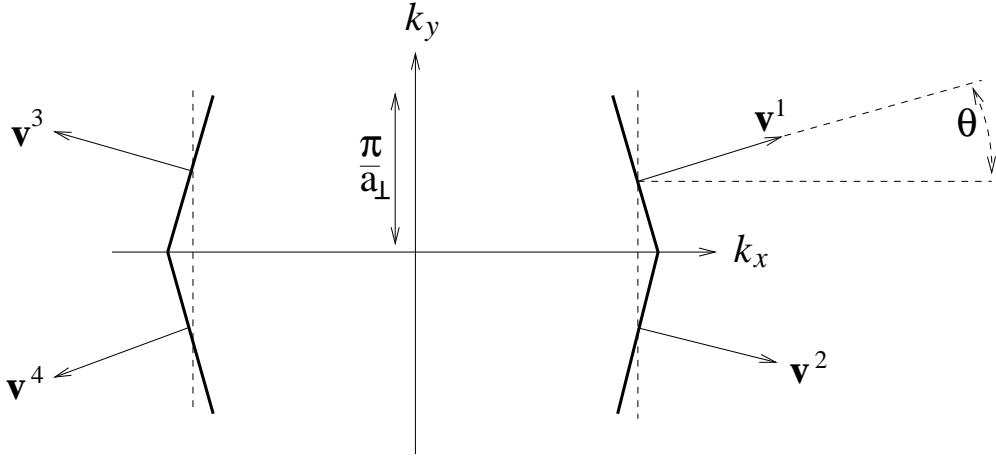}
\caption[Fermi surface of the $4$-patch model.]
{\small 
Fermi surface\index{four-patch model!Fermi surface} of the $4$-patch model. Only the intersection
with a plane of constant $k_{z}$ is shown. }
\vspace{1.0cm}
\label{fig:4patch}
\end{figure}

We now confirm the general results discussed in the previous  subsection
by explicit calculations for $M=4$, where the collective mode equation 
is quadratic and can be solved exactly.
Let us start with the 
Fermi surface shown in Fig.\ \ref{fig:4patch}.
The local Fermi velocities are
 \begin{equation}
 \begin{array}{ccl}
 {\bf{v}}^{1} & = &  ( {\bf{e}}_{x} \cos \theta + {\bf{e}}_{y} \sin \theta ) v_{F} \\
 {\bf{v}}^{2} & = & ( {\bf{e}}_{x} \cos \theta - {\bf{e}}_{y} \sin \theta ) v_{F}
 \\
 {\bf{v}}^{3} & = & ( - {\bf{e}}_{x} \cos \theta + {\bf{e}}_{y} \sin \theta ) v_{F}
 \\
 {\bf{v}}^{4} & = & ( - {\bf{e}}_{x} \cos \theta - {\bf{e}}_{y} \sin \theta ) v_{F}
 \end{array}
 \; \; \; ,
 \label{eq:velocities4patch}
 \end{equation}
and the non-interacting polarization is now
 \begin{equation}
 \Pi_{0} ( q )  
 = \frac{ \nu}{2} \sum_{\alpha = 1}^{2} 
 \frac{ ( {\bf{v}}^{\alpha} \cdot {\bf{q}})^2 }
 { ( {\bf{v}}^{\alpha} \cdot {\bf{q}})^2 + \omega_{m}^2 }
 \label{eq:Pi04patch}
 \; \; \; .
 \end{equation}
The collective mode Eq.\  (\ref{eq:PolynomM})
reduces to the bi-quadratic equation
 \begin{equation}
  z^4 - 
 \left( 1 + \frac{F_{\bf{q}} }{2} \right) 
 \left( \xi_{\bf{q}}^2 + \tilde{\xi}_{\bf{q}}^2 \right)
  z^2
 + 
  ( 1 + F_{\bf{q}} )  \xi_{\bf{q}}^2 \tilde{\xi}_{\bf{q}}^2
  = 0
 \; \; \; ,
 \label{eq:biquad4patch}
 \end{equation}
where we have introduced the notation
 \begin{eqnarray}
 \xi_{\bf{q}} & = & {\bf{v}}^{1} \cdot {\bf{q}} = v_{F} ( q_{x} \cos \theta + q_{y} \sin \theta )
 \; \; \; ,
 \label{eq:xi4patchdef}
 \\
 \tilde{\xi}_{\bf{q}} & = & {\bf{v}}^{2} \cdot {\bf{q}} = v_{F} ( q_{x} \cos \theta - q_{y} \sin \theta )
 \label{eq:barxi4patchdef}
 \; \; \; .
 \end{eqnarray}
Eq.\ (\ref{eq:biquad4patch}) is easily solved,
 \begin{eqnarray}
   \hspace{-5mm}
   \omega_{\bf{q}}^2  & = &
  \left( 1 + \frac{F_{\bf{q}}}{2} \right) 
 \frac{ \xi_{\bf{q}}^2 + \tilde{\xi}_{\bf{q}}^2 }{2}
 + \frac{1}{2}
 \left[  F_{\bf{q}}^2
 \left( \frac{ \xi_{\bf{q}}^2 + \tilde{\xi}_{\bf{q}}^2 }{2} \right)^2 +
 ( 1 + F_{\bf{q}} ) ( \xi_{\bf{q}}^2 - \tilde{\xi}_{\bf{q}}^2 )^2 \right]^{1/2}
 \; ,
 \label{eq:omega4patch}
 \\
   \hspace{-5mm}
   \tilde{\omega}_{\bf{q}}^2  & = &
  \left( 1 + \frac{F_{\bf{q}}}{2} \right) 
 \frac{ \xi_{\bf{q}}^2 + \tilde{\xi}_{\bf{q}}^2 }{2}
 - \frac{1}{2}
 \left[  F_{\bf{q}}^2
 \left( \frac{ \xi_{\bf{q}}^2 + \tilde{\xi}_{\bf{q}}^2 }{2} \right)^2 +
 ( 1 + F_{\bf{q}} ) ( \xi_{\bf{q}}^2 - \tilde{\xi}_{\bf{q}}^2 )^2 \right]^{1/2}
 \;  .
 \label{eq:baromega4patch}
 \end{eqnarray}
The right-hand side of Eq.\ (\ref{eq:baromega4patch}) is non-negative because
 \begin{eqnarray}
  \left[ \left( 1 + \frac{F_{\bf{q}}}{2} \right) 
 \frac{ \xi_{\bf{q}}^2 + \tilde{\xi}_{\bf{q}}^2 }{2} \right]^2
 - \frac{1}{4}
 \left[  F_{\bf{q}}^2
 \left( \frac{ \xi_{\bf{q}}^2 + \tilde{\xi}_{\bf{q}}^2 }{2} \right)^2 +
 ( 1 + F_{\bf{q}} ) ( \xi_{\bf{q}}^2 - \tilde{\xi}_{\bf{q}}^2 )^2 \right]
 & & 
 \nonumber
 \\
 & & \hspace{-110mm} =  ( 1 + F_{\bf{q}} ) \xi_{\bf{q}}^2 \tilde{\xi}_{\bf{q}}^2 \geq 0
 \label{eq:positivecheck}
 \; \; \; .
 \end{eqnarray}
Both modes $\omega_{\bf{q}}$ and $\tilde{\omega}_{\bf{q}}$ give
rise to $\delta$-function peaks in the dynamic structure factor.
We obtain
for $\omega > 0$ 
 \begin{equation}
 S^{RPA} ( {\bf{q}} , \omega ) = Z_{\bf{q}} \delta ( \omega - \omega_{\bf{q}} )
 + \tilde{Z}_{\bf{q}} \delta ( \omega - \tilde{\omega}_{\bf{q}} )
 \label{eq:SRPA4patch}
 \; \; \; ,
 \end{equation}
with the residues given by
 \begin{eqnarray}
 Z_{\bf{q}} & = & \frac{\nu}{2 \omega_{\bf{q}} }
 \frac{ 
 \omega_{\bf{q}}^2 \frac{ \xi_{\bf{q}}^2 + \tilde{\xi}_{\bf{q}}^2 }{2}  - \xi_{\bf{q}}^2
 \tilde{\xi}_{\bf{q}}^2 }{ \omega_{\bf{q}}^2 - \tilde{\omega}_{\bf{q}}^2 }
 \; \; \; ,
 \label{eq:Z1patch}
 \\
 \tilde{Z}_{\bf{q}} & = &  \frac{\nu}{2 \tilde{\omega}_{\bf{q}} }
 \frac{ 
 \xi_{\bf{q}}^2
 \tilde{\xi}_{\bf{q}}^2 
 - \tilde{\omega}_{\bf{q}}^2 \frac{ \xi_{\bf{q}}^2 + \tilde{\xi}_{\bf{q}}^2 }{2}   }
 { \omega_{\bf{q}}^2 - \tilde{\omega}_{\bf{q}}^2 }
 \label{eq:Z2patch}
 \; \; \; .
 \end{eqnarray}
In the limit $\theta \rightarrow 0$ we have $\tilde{\xi}_{\bf{q}} \rightarrow \xi_{\bf{q}}
\rightarrow v_{F} q_{x}$, so that
$\omega_{\bf{q}} \rightarrow \sqrt{ 1 + F_{\bf{q}} } v_{F} | q_{x} |$ and
$\tilde{\omega}_{\bf{q}} \rightarrow v_{F} | q_{x} |$.
It is also easy to see that the residue $Z_{\bf{q}}$ in this limit reduces
to the result given in Eq.\ (\ref{eq:Zom1patch}), while the residue
$\tilde{Z}_{\bf{q}}$ vanishes. 
To examine whether interchain hopping destroys
the Luttinger liquid behavior, it is sufficient
to calculate the static Debye-Waller factor.
Combining Eqs.\ (\ref{eq:Rlondef}), (\ref{eq:Slondef}) and (\ref{eq:SRPA4patch}),
it is easy to show that 
 \begin{eqnarray}
 Q^{\alpha} ( r_{\|} \hat{\bf{v}}^{\alpha} , 0 )
 & = & 
 - \frac{1}{V} \sum_{\bf{q}} 
 \left[ 1 - \cos ( {\hat{\bf{v}}^{\alpha} } \cdot {\bf{q}} r_{\|} ) \right]
 f_{\bf{q}}^2 \int_{0}^{\infty} d \omega
 \frac{S^{RPA} ( {\bf{q}} , \omega )}{ ( \omega + | {\bf{v}}^{\alpha} \cdot
 {\bf{q}} |)^2 }
 \nonumber
 \\ 
 &= & 
 - \frac{1}{V} \sum_{\bf{q}} 
 \left[ 1 - \cos ( {\hat{\bf{v}}^{\alpha} } \cdot {\bf{q}} r_{\|} ) \right]
 f_{\bf{q}}^2 
 \left[ \frac{Z_{\bf{q}}}{ ( \omega_{\bf{q}} + 
 | {\bf{v}}^{\alpha} \cdot {\bf{q}} |)^2}
 +  \frac{\tilde{Z}_{\bf{q}}}{ ( \tilde{\omega}_{\bf{q}} + 
 | {\bf{v}}^{\alpha} \cdot {\bf{q}} |)^2}
 \right]
 \label{eq:Q4patch1}
 \; \; \; .
 \end{eqnarray}

\subsubsection{The plasmon mode}

Let us first focus on the first term in the second line
of Eq.\ (\ref{eq:Q4patch1}). Because for $t_{\bot} \rightarrow 0$
this term smoothly reduces to the corresponding expression 
in the absence of interchain hopping, the mode $\omega_{\bf{q}}$ can
be identified with the physical plasmon mode discussed in
Sec.\ \ref{subsec:Mpatch}.
From Sec.\ \ref{sec:thecoulomb} we know 
that the Debye-Waller factor is essentially determined by the
long wavelength regime $| {\bf{q}} |
{ \raisebox{-0.5ex}{$\; \stackrel{<}{\sim} \;$}} \kappa$. 
In this regime $F_{\bf{q}} \gg 1$, so that we may 
use the strong-coupling approximation for
$\omega_{\bf{q}}$ and $Z_{\bf{q}}$.
From Eqs.\ (\ref{eq:omega4patch}) and (\ref{eq:baromega4patch}) it is easy to see that,
up to higher orders in $\theta / F_{\bf{q}}$, the collective density mode
$\omega_{\bf{q}}$ can in this regime be approximated by
 \begin{eqnarray}
 \omega_{\bf{q}} & \approx & v_{F} \sqrt{ 1 + F_{\bf{q}} }  \sqrt{ q_{x}^2 + \theta^2 q_{y}^2 }
 \; \; \; ,
 \label{eq:om4SC}
 \; \; \; .
 \end{eqnarray}
Substituting this expression into Eq.\ (\ref{eq:Z1patch}), we obtain
 \begin{eqnarray}
 Z_{\bf{q}} & \approx & \frac{\nu v_{F} \sqrt{ q_{x}^2 + \theta^2 q_{y}^2}}{2 \sqrt{1+F_{\bf{q}}} }
 \label{eq:Z1patchSC}
 \; \; \; .
 \end{eqnarray}
Comparing Eq.\ (\ref{eq:Z1patchSC}) with Eq.\ (\ref{eq:Zom1patch}), we see that
the only effect of the interchain hopping is the replacement
$| q_{x} | \rightarrow \sqrt{ q_{x}^2 + \theta^2 q_{y}^2 }$.
Thus, the plasmon mode $\omega_{\bf{q}}$ yields
the following contribution
to the constant part
of Eq.\ (\ref{eq:Q4patch1})
for small $\theta$
 \begin{equation}
 R^{\alpha}_{pl} =  - \frac{1}{2} \int_{0}^{ \infty } d q_{x}
 \left< 
 \frac{1}{ \sqrt{ q_{x}^2 + \theta^2 q_{y}^2 }} 
  \frac{F_{\bf{q}}^2 }
 { 2 \sqrt{ 1 + F_{\bf{q}} } 
 \left[ \sqrt{ 1 + F_{\bf{q}} } + 1
 \right]^2 }
 \right>_{BZ}
 \label{eq:R4patch2}
 \; \; \; .
 \end{equation}
If we set $\theta = 0$ we recover the previous result
in the absence of interchain hopping (Eq.\ (\ref{eq:R3patch})), which is
logarithmically divergent. This divergence is due to the fact
that for $\theta = 0$ the first factor in Eq.\ (\ref{eq:R4patch2})
can be pulled out of the averaging bracket. However, for any
finite $\theta$ the $q_{x}$- and $q_{y}$-integrations are coupled, 
so that it is not possible to factorize the integrations. Hence,
any non-zero value of $\theta$ couples the phase space 
of the ${\bf{q}}$-integration.
Because for $\theta \rightarrow 0$ the integral in Eq.\ (\ref{eq:R4patch2}) is 
logarithmically
divergent, the coefficient of the leading logarithmic 
term can be extracted by ignoring the $q_{x}$-dependence of
the second factor in the averaging symbol of Eq.\ (\ref{eq:R4patch2}). 
Then we obtain to leading logarithmic order
 \begin{eqnarray}
 R^{\alpha}_{pl} & \sim &  - \frac{1}{2} \int_{0}^{ \kappa } d q_{x}
 \left< 
 \frac{1}{ \sqrt{ q_{x}^2 + \theta^2 q_{y}^2 }} 
 \lim_{q_{x} \rightarrow 0}
 \left[ \frac{F_{\bf{q}}^2 }
 { 2 \sqrt{ 1 + F_{\bf{q}} } 
 \left[ \sqrt{ 1 + F_{\bf{q}} } + 1
  \right]^2 }
 \right]
 \right>_{BZ}
 \nonumber
 \\
 & = & - \frac{1}{2} \left< 
 \ln \left( \frac{\kappa}{\theta | q_{y} | } \right)
 \lim_{q_{x} \rightarrow 0}
 \left[ \frac{F_{\bf{q}}^2 }
 { 2 \sqrt{ 1 + F_{\bf{q}} } 
 \left[ \sqrt{ 1 + F_{\bf{q}} } + 1
 \right]^2 }
 \right] \right>_{BZ}
 \nonumber
 \\
 & = &
 - \gamma_{cb} \left[ \ln \left( \frac{1}{\theta} \right) + b_{1} \right]
 \label{eq:Rres4patch}
 \; \; \; ,
 \end{eqnarray}
where $\gamma_{cb}$ is given in Eq.\ (\ref{eq:gammacbres}), and
$b_{1}$ is a numerical constant of the order of unity.

The contribution of the plasmon mode to the
spatially varying part 
$S^{\alpha} ( r_{\|} \hat{\bf{v}}^{\alpha} , 0 )$
of the Debye-Waller factor at equal times 
can be calculated analogously.
Note that
$ r_{\|} = \hat{\bf{v}}^{\alpha} \cdot {\bf{r}}
= \pm r_{x} \cos \theta  \pm r_{y} \sin \theta$.
Repeating the steps leading to Eq.\ (\ref{eq:R4patch2}), we obtain
 \begin{eqnarray}
 S^{\alpha}_{pl}  ( r_{\|} \hat{\bf{v}}^{\alpha} , 0 ) & = & - 
 \frac{1}{2} \int_{0}^{ \infty } d q_{x}
 \cos ( q_{x} r_{\|} )
 \left< 
 \frac{ \cos ( \theta q_{y} r_{\|})}{ \sqrt{ q_{x}^2 + \theta^2 q_{y}^2 }} 
  \frac{F_{\bf{q}}^2 }
 { 2 \sqrt{ 1 + F_{\bf{q}} } 
 \left[ \sqrt{ 1 + F_{\bf{q}} } + 1
  \right]^2 }
 \right>_{BZ}
 \; \; \; .
 \nonumber
 \\
 & &
 \label{eq:S4patch2}
 \end{eqnarray}
Because the Thomas-Fermi wave-vector $\kappa$ acts as an effective
ultraviolet cutoff, 
the value of the integral in Eq.\ (\ref{eq:S4patch2}) is determined
by the regime $|{\bf{q}} | 
{ \raisebox{-0.5ex}{$\; \stackrel{<}{\sim} \;$}}
\kappa$. 
For $\theta \kappa | r_{\|} | \ll 1$ we may approximate
in this regime
 $ \cos (\theta q_{y} r_{\|}) \approx 1$ under the integral sign. 
Furthermore, for $\kappa | r_{\|}| \gg 1$ the oscillating 
factor $\cos ( q_{x} r_{\|} )$ effectively replaces $\kappa$ by
$|r_{\|}|^{-1}$ as the relevant ultraviolet cutoff. 
We conclude that  in the  parametrically large  intermediate regime 
 \begin{equation}
 \kappa^{-1} \ll | r_{\|} | \ll  ( \theta \kappa )^{-1}
 \label{eq:regimeinterm}
 \; \; \; ,
 \end{equation}
we have 
to leading logarithmic order 
 \begin{equation}
 S^{\alpha}_{pl} ( r_{\|} \hat{\bf{v}}^{\alpha} , 0 ) \sim
 - \gamma_{cb} \left[ \ln \left( \frac{1}{\theta \kappa | r_{\|} | } \right) + b_{2} \right]
 \; \; \; ,
 \label{eq:Sres4patch}
 \end{equation}
where $b_{2}$ is another numerical constant.

\subsubsection{The nesting mode}
\label{subsubsec:nesting}

\begin{figure}[hit]
\vspace{1.0cm}
\hspace{1cm}
\epsfysize6cm
\epsfbox{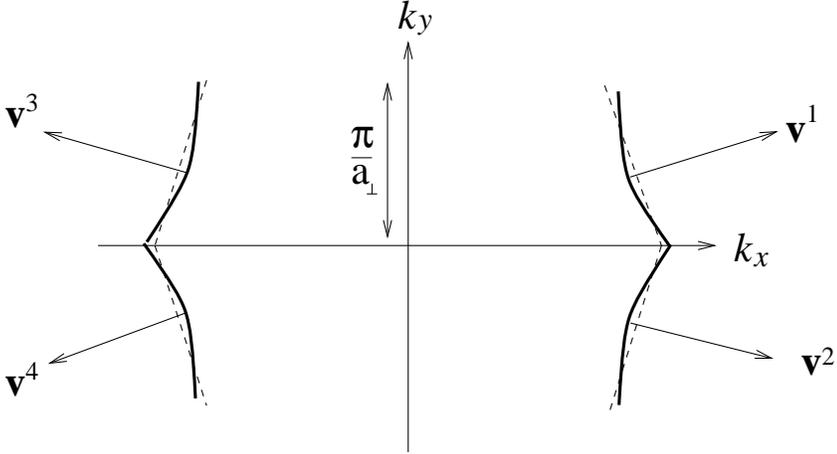}
\caption[Fermi surface of the $4$-patch model with curved patches]
{\small 
Fermi surface\index{four-patch model!Fermi surface} of the $4$-patch model
with curved patches. 
If the component of ${\bf{q}}$ perpendicular to ${\bf{v}}^{\alpha}$
is denoted by $q_{\bot}$,
the patches can be described by energy dispersions
$\xi^{\alpha}_{\bf{q}} = {\bf{v}}^{\alpha} \cdot {\bf{q}} +
\frac{ q_{\bot}^2 }{ 2 m_{\bot}}$ with negative effective mass $m_{\bot}$.
}
\vspace{1.0cm}
\label{fig:4patchcurved}
\end{figure}

From the general analysis presented in Sec.\ \ref{subsec:Mpatch}
we expect that the second mode $\tilde{\omega}_{\bf{q}}$ in
Eq.\ (\ref{eq:Q4patch1}) will give rise to an  unphysical 
logarithmic growth of
$Q^{\alpha} ( r_{\|} \hat{\bf{v}}^{\alpha} , 0 )$ at large distances, 
which is caused
by the nesting symmetry of the Fermi surface in our simple
$4$-patch model.
We now verify this expectation 
and then use the results of Refs.\ \cite{Kopietzhab,Kopietz96} to show
by explicit calculation that curvature 
effects remove the nesting singularities.
For convenience
we choose the integration variables 
$q_{\|} = {\hat{\bf{v}}}^{1} \cdot {\bf{q}} = q_x \cos \theta + q_y \sin \theta $ and
$q_{\bot} = - q_x \sin \theta + q_y \cos \theta$. 
Then $\xi_{\bf{q}} = v_F q_{\|}$ and 
$\tilde{\xi}_{\bf{q}} = v_F ( q_{\|} - 2 \theta q_{\bot} )$ to leading order
in $\theta$.
Note that the condition
$\mbox{$ | q_{\|} | 
{ \raisebox{-0.5ex}{$\; \stackrel{<}{\sim} \;$}}
\theta | q_{\bot} | $}$
is equivalent with $| \xi_{\bf{q}} | 
{ \raisebox{-0.5ex}{$\; \stackrel{<}{\sim} \;$}}
| \tilde{\xi}_{\bf{q}} |$.
Geometrically this means that the wave-vector ${\bf{q}}$ is almost
parallel to the surface of the first and fourth patch, so that
its projection $ \hat{\bf{v}}^{1} \cdot {\bf{q}} =  
- \hat{\bf{v}}^{4} \cdot {\bf{q}}$ 
on the local normals
is much smaller than the projection 
on the normals $\hat{\bf{v}}^{2}$ 
and $\hat{\bf{v}}^{3}$ 
of the other two patches.
In this regime
we obtain from Eq.\ (\ref{eq:baromega4patch}) to leading order
 \begin{equation}
   \tilde{\omega}_{\bf{q}}  \approx
 \sqrt{\frac{  1 + F_{\bf{q}}  }{ 1 +  \frac{F_{\bf{q}}}{2} }}
 v_{F} | q_{\|} |
 \; \; \; , \; \; \; | q_{\|} | 
{ \raisebox{-0.5ex}{$\; \stackrel{<}{\sim} \;$}}
 \theta | q_{\bot} |
 \; \; \; ,
 \label{eq:omegabarcrit}
 \end{equation}
and from Eq.\ (\ref{eq:Z2patch}) 
 \begin{equation}
 \tilde{Z}_{\bf{q}} \approx
 \frac{ \nu v_F | q_{\|} |}{ 4   
 [ 1 + \frac{F_{\bf{q}}}{2} ]^{\frac{3}{2} }
 [ 1 + F_{\bf{q}} ]^{\frac{1}{2}} }
 \; \; \; , \; \; \; | q_{\|} | 
{ \raisebox{-0.5ex}{$\; \stackrel{<}{\sim} \;$}}
 \theta | q_{\bot} |
 \label{eq:Z2patchcrit}
 \; \; \; .
 \end{equation}
We conclude that
for
 $ | q_{\|} | 
{ \raisebox{-0.5ex}{$\; \stackrel{<}{\sim} \;$}}
 \theta | q_{\bot} |$
 \begin{equation}
  \frac{\tilde{Z}_{\bf{q}}}{ ( \tilde{\omega}_{\bf{q}} + 
 | {\bf{v}}^{\alpha} \cdot {\bf{q}} |)^2}
 \approx
 \frac{ \nu }{
 4 v_F | q_{\|} | 
 [ 1 + \frac{F_{\bf{q}}}{2} ]^{\frac{3}{2} }
 [ 1 + F_{\bf{q}} ]^{\frac{1}{2}} 
 [ \sqrt{ \frac{ 1 + F_{\bf{q}} }{ 1 + \frac{ F_{\bf{q}}}{2} } }  + 1 ]^2}
 \;  \; \; .
  \label{eq:nestsing2}
  \end{equation}
Note that 
the  above expressions agree with Eqs.\ (\ref{eq:nestM})--(\ref{eq:Rnest}) 
if we set there $M = 4$.
The factor of $ 1 / | q_{\|} |$ in Eq.\ (\ref{eq:nestsing2})
implies that the static Debye-Waller factor grows logarithmically
for  $r_{\|} \rightarrow \infty$.
However, for realistic Fermi surfaces of the
form shown in Fig.\ \ref{fig:quasi1d} the nesting symmetry 
responsible for this behavior does not exist.
To remove this artificial nesting symmetry, let us now replace the
completely flat patches of Fig.\ \ref{fig:4patch} by
the slightly curved patches
shown in Fig.\ \ref{fig:4patchcurved}.
The corresponding energy dispersions 
can be taken to be
 $\xi^{\alpha}_{\bf{q}} = {\bf{v}}^{\alpha} \cdot {\bf{q}} 
 + \frac{  q_{\bot}^2 }{2  m_{\bot}  }$
with negative effective mass $m_{\bot}$. 
For weakly coupled chains we should choose
$|m_{\bot}| \approx  m_{\|} / \theta$, where $m_{\|} = k_F / v_F$ is the
effective mass for motion along the chains.
As shown in Refs.\ \cite{Kopietzhab,Kopietz96}, one important
effect of the curvature terms 
on the Debye-Waller factor is the
replacement given in Eq.\ (\ref{eq:polereplace}), 
which removes the double pole in Eqs.\ (\ref{eq:Rlondef}) and (\ref{eq:Slondef}).
Taking this effect into account, we see that
the contribution of the nesting mode to the constant part
$R^{\alpha}$ of Eq.\ (\ref{eq:Q4patch1})
becomes
 \begin{equation}
 R^{\alpha}_{nest} \approx 
   -  
 \frac{1}{V} \sum_{\bf{q}} 
 f_{\bf{q}}^2
 \tilde{Z}_{\bf{q}}
 \frac{ 2 \mbox{sgn} ( \xi^{\alpha}_{\bf{q}}  ) }
 {  \frac{ q_{\bot}^2 }{ |m_{\bot}|  }  
 ( \tilde{\omega}_{\bf{q}} + | \xi^{\alpha}_{\bf{q}} |) }
 \label{eq:R2nest}
 \; \; \; ,
 \end{equation}
with $\tilde{Z}_{\bf{q}}$ and $\tilde{\omega}_{\bf{q}}$ given in
Eqs.\ (\ref{eq:Z2patch}) and (\ref{eq:baromega4patch}).
From the above discussion  we know that
possible nesting singularities 
are due to the regime $| q_{\|} | 
{ \raisebox{-0.5ex}{$\; \stackrel{<}{\sim} \;$}}
   \theta | q_{\bot} |$.
Thus, restricting the limits of the integral in Eq.\ (\ref{eq:R2nest}) to this
regime, we obtain 
from Eqs.\ (\ref{eq:baromega4patch}) and (\ref{eq:Z2patch})
in the strong coupling limit
 \begin{eqnarray}
 f_{\bf{q}}^2
 \tilde{Z}_{\bf{q}} & \approx & \frac{v_F | q_{\|} |}{\sqrt{2} \nu}
 \label{eq:fZbarstrong}
 \; \; \; ,
 \\
 \tilde{\omega}_{\bf{q}} & \approx & \sqrt{2} v_F | q_{\|} |
 \label{eq:omegabarstrong}
 \; \; \; .
 \end{eqnarray}
Recall that for the three-dimensional Coulomb interaction
the strong-coupling condition $\nu f_{\bf{q}} \gg 1$
is equivalent with $| {\bf{q}} | \ll \kappa$. 
Putting everything together, we 
find that the contribution from the
critical regime $| q_{\|} | 
{ \raisebox{-0.5ex}{$\; \stackrel{<}{\sim} \;$}}
\theta | q_{\bot} |$ to
Eq.\ (\ref{eq:R2nest}) can be written as
 \begin{equation}
 R^{\alpha}_{nest} \approx 
 - \frac{\sqrt{2} \kappa }{\pi^3 \nu} \int_{0}^{\kappa}
 d q_{\bot} 
 \frac{ | m_{\bot} | }{ q_{\bot}^2 }
 \int_{- \theta | q_{\bot} |}^{\theta | q_{\bot} | } d q_{\|}
 \frac{ 
  | q_{\|} | \mbox{sgn} ( q_{\|} - \frac{ q_{\bot}^2}{2 
 | m_{\bot}| v_{F} } ) }{  \sqrt{2} | q_{\|} | + | q_{\|} -
 \frac{ q_{\bot}^2}{ 2 | m_{\bot} | v_{F}  } |  }
 \label{eq:R2nestcrit}
 \; \; \; .
 \end{equation}
The $q_{\|}$-integration can now be performed analytically.
The integral is proportional to
$\frac{ q_{\bot}^2}{  |m_{\bot}| }$, which cancels 
the singular factor of $\frac{ | m_{\bot} | }{ q_{\bot}^2 }$ 
in Eq.\ (\ref{eq:R2nestcrit}).
We obtain
\begin{equation}
 R^{\alpha}_{nest} \approx 
 - \frac{\kappa }{ \sqrt{2} ( \sqrt{2} + 1 )^2 \pi^3 \nu v_F} 
 \int_{0}^{\kappa} d q_{\bot} \left[ \ln \left( \frac{ 2 |m_{\bot}| v_F 
 \theta}{ q_{\bot} } \right) + b_3 \right]
 \; \; \; ,
 \label{eq:R2nestcrit2}
 \end{equation}
where $b_3$ is a numerical constant of the order of unity.
This is the same type of 
integral as in Eq.\ (\ref{eq:curvaturesub}), so that 
our simple intuitive arguments given at the end of 
Sec.\ \ref{subsec:Mpatch}
are now put on a more solid basis.
As already mentioned, in physically relevant cases we expect 
$| m_{\bot} | v_F \theta \approx k_F$, 
so that we finally obtain
 $R^{\alpha}_{nest} \approx 
 - \gamma_{cb} b_4$
where $b_4$ is another numerical constant of the order of unity.
Thus, for patches with finite curvature the contribution
of the nesting mode is finite. 
It is also easy to see that the curvature terms do
{\it{not}} modify the logarithmic small-$\theta$ behavior
of $R^{\alpha}$ due to the plasmon mode given in Eq.\ (\ref{eq:Rres4patch}). 
This is so because
the leading $\ln ( {1}/{\theta} )$-term in Eq.\ (\ref{eq:Rres4patch})
is generated by the energy scale $v_F \theta | q_{\bot} |$, which is
by assumption larger than the curvature energy $ q_{\bot}^2/( 2 | m_{\bot} | )$.
Comparing our result for $R^{\alpha}_{nest}$ with the
contribution from the plasmon mode given in
Eq.\ (\ref{eq:Rres4patch}), we see that for small $\theta$
the contribution of the nesting mode 
does not modify the leading logarithmic behavior
$R^{\alpha}_{pl} \sim - \gamma_{cb} \ln ( {1}/{\theta} )$ due to the 
plasmon mode.
Hence, to leading logarithmic order, 
the static Debye-Waller factor is dominated by the
contribution from the plasmon mode, so that we may write
$R^{\alpha} \approx R^{\alpha}_{pl}$, and similarly for $S^{\alpha} ( r_{\|}
\hat{\bf{v}}^{\alpha} , 0 )$.

\subsubsection{Anomalous scaling}
\label{subsubsec:anomalscale}

Because $R^{\alpha}$ is finite 
for any non-zero $\theta$,
the system is a Fermi liquid, 
with quasi-particle residue
 \begin{equation}
 Z^{\alpha} = e^{R^{\alpha}} \propto \theta^{\gamma_{cb}}
 \label{eq:QPres4patch}
 \; \; \; .
 \end{equation}
Thus, for $\theta \rightarrow 0$ the quasi-particle residue
vanishes with a non-universal power of $\theta$, which 
{\it{can be identified with the anomalous dimension of the corresponding
Luttinger liquid that would exist for $\theta = 0$ at the same value of the
dimensionless coupling constant $g $.}}
Combining Eqs.\ (\ref{eq:Rres4patch}) and
(\ref{eq:Sres4patch}), we obtain for the total static Debye-Waller factor
 \begin{eqnarray}
 Q^{\alpha} ( r_{\|} \hat{\bf{v}}^{\alpha} , 0 ) & = &
 R^{\alpha} -
 S^{\alpha} ( r_{\|} \hat{\bf{v}}^{\alpha} , 0 )
 \nonumber
 \\
 & = & - \gamma_{cb} 
 \left[ \ln ( \kappa | r_{\|}| ) + b  \right]
 \; \; \; ,
 \; \; \; \mbox{ $ \kappa^{-1}  \ll | r_{\|} | \ll  ( \theta \kappa )^{-1}$ }
 \; \; \; ,
 \label{eq:Qres4patch}
 \end{eqnarray}
where $b$ is a numerical constant of the order of unity.
Exponentiating this expression, we see that the
interacting Green's function satisfies the anomalous scaling 
relation,
 \begin{equation}
 G^{\alpha} ( {\bf{r}} / s  ,  0 ) = s^{3 + \gamma_{cb}}
 G^{\alpha} ( {\bf{r}}  ,  0 )
 \; \; \; ,
 \; \; \; \mbox{ $\kappa^{-1}  \ll | r_{\|} | \; , \; |r_{\|}| / s   \ll  (\theta \kappa )^{-1}$ }
 \; \; \; .
 \end{equation}
In momentum space this implies for $\omega = 0$ and 
$ \theta \kappa \ll | {\bf{q}} | \ll \kappa$ an anomalous scaling law of the
form given in Eq.\ (\ref{eq:anomalscale}) with 
$\gamma = \gamma_{cb}$.
Thus, in spite of the fact that the system is a Fermi liquid, there exists for small
$\theta$ a parametrically large intermediate regime where the interacting Green's function
satisfies the  anomalous scaling law of Luttinger liquids.
Moreover, the effective anomalous exponent 
{\it{is precisely given by the
anomalous dimension of the Luttinger liquid that would exist for $\theta = 0$}}.
This is a very important result, 
because in realistic experimental systems the interchain hopping $t_{\bot}$ can 
never be completely turned off.
Our result implies that for small but finite $\theta$ the
anomalous dimension of the Luttinger liquid is in principle measurable,
although strictly speaking the system is a Fermi liquid.

\section{Comparison with other methods}
\label{sec:wenapproximation}

As mentioned in the introduction, an alternative way to
approach the problem in the limit of weak interchain hopping
is by straight or renormalization group aided perturbation theory in
$t_{\bot}$\cite{Wen90,Bourbonnais91,Castellani92,Boies95,Clarke96,Tsvelik96}.
In contrast to our result using higher dimensional bosonization
these methods yield a finite value for the quasi-particle weight
only for values of the anomalous dimensions smaller than
a critical value.

Starting point in the approximation proposed by
Wen\cite{Wen90} and elaborated on by others
\cite{Bourbonnais91,Boies95,Clarke96,Tsvelik96}
is to expand the self-energy defined in the
usual way by
\begin{equation}
 G ( {\bf{k}} , \omega ) = 
 \frac{1}{ \omega - \epsilon_{\bf{k}} - \Sigma ( {\bf{k}} , \omega ) }
 \label{eq:GDyson}
 \end{equation}
in powers of the hopping. If one defines $\delta \Sigma ( {\bf{k}} , \omega )
\equiv \Sigma ( {\bf{k}} , \omega ) 
- \left. \Sigma ( {\bf{k}} , \omega )  \right|_{t_{\bot} = 0 }$ and linearizes the
dispersion in the direction of the chains near $k_F$, one obtains
for the model where only intrachain electron-electron interactions are taken
into account
\begin{equation}
 G ( {\bf{k}} , \omega ) = \frac{1}{ 
 g ( k_{\|} , \omega )^{-1} - t_{\bot}  ( {\bf{k}}_{\bot} ) 
 - \delta \Sigma ( {\bf{k}} , \omega ) }
 \; \; \; ,
 \label{eq:GWen}
 \end{equation}
where $g ( k_{\|} , \omega )$ is the exact interacting single chain
propagator. 
Wen\cite{Wen90} and later others
\cite{Bourbonnais91,Boies95}
then introduce the rather crude approximation to
neglect $\delta \Sigma$ completely.
This is justified by Wen by the fact that
$\delta \Sigma$ formally vanishes with a higher power in $t_{\bot}$ than the
linear term kept. For the system to be a Fermi liquid a zero frequency
pole should occur. This can happen for arbitrarily small hopping
only if $ g ( k_{\|} , 0 )$ diverges. Using the exact results
for the one-dimensional Green's function, $g ( k_{\|} , 0 )$
can in fact be shown to diverge (for the model without and with spin)
for values of the anomalous dimension $\gamma < 1$.
In this case the quasi-particle weight in this approximation
is proportional to $ ( t_{\bot} / E_F )^{ \gamma / ( 1 - \gamma )}$, and therefore
{\it{vanishes}} when $\gamma$ reaches one. This result is in agreement
with the naive renormalization group argument in which the weak-coupling
criterion for the relevance of $t_{\bot}$ is used for
arbitrary strong coupling. This behavior of the quasi-particle weight differs
from our bosonization result, which yields a finite value also 
for $\gamma = 1$ and larger.
The shape of the Fermi surface for $\gamma < 1$ in Wen's approximation
is determined by the equation $g ( k_{\|} , 0 ) = 1 / 
t_{\bot} ( {\bf{k}}_{\bot} )$
and yields a modification $\delta k_F / k_F$ compared to the flat
surface of zero transverse hopping of order $(t_{\bot} / E_F )^{1 / ( 1 - \gamma)}$,
i.e. the {\it{shape}} of the Fermi surface depends on the {\it{strength}} of the
electron-electron interaction. Such a dependence is
missing in our bosonization approach. This is probably
the reason for the qualitatively different results for the
quasi-particle weight for $\gamma \approx 1$.

If one calculates the spectral function corresponding to Wen's Green's function,
it shows various properties in the low energy regime which look
rather unphysical.
If one crosses the Fermi surface  there are sharp poles even for
finite distances away from it on one side, but there is only
continuous spectral weight on the other side. We believe
our bosonization result to produce
more reliable spectra, at least for $\gamma \ll 1$. For a
direct comparison for our model
including electron-electron interchain interaction,
it would be necessary to first generalize Wen's approximation
to this case. This can be done be replacing $g ( k_{\|} , \omega )$
in Eq.(\ref{eq:GWen}) by the exact Green's function for zero
hopping which takes into account the interchain interaction.
For this model the approximation to neglect $\delta \Sigma$ completely
is even more serious, as $\delta \Sigma$ vanishes with one power in
$t_{\bot}$ less than in the model with intrachain 
electron-electron interaction only. This can be easily shown
diagrammatically.

Clarke and Strong\cite{Clarke96} have argued that even in the range
$1/2 < \gamma < 1$ the Green's function in Wen's approximation
contains unphysical non-analyticities which indicate a breakdown of Fermi liquid
behavior already at $\gamma = 1/2$. This seems to fit well to the
other results from their concept of ``confined coherence''. It will
be shown elsewhere\cite{Schonhammer97} that their arguments
concerning the analytical properties of Wen's Green's function are not
well justified.

Even if one assumes it to be correct that the Fermi surface becomes
flat as $\gamma$ approaches one (from below), the resulting
state of the system cannot be a simple Luttinger liquid, defined
as a non-trivial metallic ground state where different
instabilities mutually cancel. Using the parquet approach it has
recently been shown\cite{Zheleznyak96} that two-dimensional systems
with flat regions on opposite sides of the Fermi surface
always develop some sort of instability towards a phase with
spontaneously broken symmetry.

\section{Conclusions}
\label{sec:conclusions}

In this work we have used the 
higher-dimensional bosonization approach to study the
problem of  coupled Luttinger liquids.
Our main result is that strictly speaking Luttinger liquid behavior 
exists only for $t_{\bot} = 0$. Any finite value of the interchain hopping
leads to a finite quasi-particle residue, which we have explicitly calculated.
Nevertheless, in a large intermediate regime of wave-vectors and frequencies
the Green's function exhibits precisely the same
scaling behavior as for $t_{\bot} = 0$. Keeping in mind that
experiments are always performed with finite resolution, the
intermediate scaling regime seems to be the experimentally relevant one --
in this sense the interchain hopping is irrelevant.

Although our approach is non-perturbative in the sense that
an infinite number of Feynman diagrams have been summed, it is
only approximate. However, 
we have a strong non-perturbative 
argument why all
terms beyond the Gaussian approximation that
have been ignored in Eqs.\ (\ref{eq:Gbosres})--(\ref{eq:Slondef})
are negligible  in the parameter regime of interest:
The closed loop theorem \cite{Kopietz94,Kopietzhab}, which is essentially equivalent
with the Ward-identity derived 
by Castellani, Di Castro and Metzner \cite{Castellani94},
guarantees a cancellation of the leading infrared singularities
in the non-Gaussian terms {\it{to all orders in perturbation theory}}.
Of course, for linearized energy 
dispersion the shape of the Fermi surface is fixed.
Therefore, if the renormalization of the shape of the Fermi surface
by the interaction becomes relevant in the present problem, 
our conclusions are expected to be modified.
However,
as discussed in Sec.\ref{sec:wenapproximation},  
at least as long as the value of the anomalous dimension
without interchain hopping is small compared with unity, 
we expect the interaction-induced modification of the shape
of the Fermi surface to be  unimportant.
It should also be mentioned that only small momentum transfers have been taken
into account in our approach, so that possible
instabilities due to $2 k_F$-processes have been neglected.
Thus, 
although we have for simplicity  taken the zero-temperature limit,
our results are implicitly restricted to temperatures $T$
where the system is in the normal state.
It is easy to see for $T  > 0$ the expressions for the
long-distance behavior of the
static Debye-Waller factor 
$Q^{\alpha} ( r_{\|} \hat{\bf{v}}^{\alpha} , 0 )$
derived above
remain correct at distances small 
with the thermal de Broglie wavelength $\lambda_{th} = h v_F / T$.
Beyond this length scale  we find
that
$Q^{\alpha} ( r_{\|} \hat{\bf{v}}^{\alpha} , 0 )$
is proportional to $- | r_{\|} | / \lambda_{th}$.
We therefore conclude that for $( \theta \kappa )^{-1} \ll \lambda_{th}$
the intermediate scaling regime discussed in
Sec.\ \ref{subsubsec:anomalscale} exists even at finite temperature.

\section*{Acknowledgments}
This research was supported in part by the National Science Foundation
under Grant No. PHY94-07194 during a stay of one of us (K. S.)
at the ITP-workshop on Non-Fermi-Liquid Behavior in Solids.
V. M. is grateful to the Deutsche Forschungsgemeinschaft and
the NSF Grant No. DMR-9416906 for financial
support.
The work of
P. K. and V. M. was partially supported by the ISI
Foundation and EU HC\&M Network ERBCHRX-CT920020.
In particular, P. K. and V. M. 
would like to thank Richard Hlubina
for stimulating discussions during a workshop at
Villa Gualino (Torino), which 
motivated us to take a closer look at the nesting problem.
We would also like to thank Steven Strong for communications.




\end{document}